\documentclass[11pt]{article}

\usepackage{xspace}
\usepackage{amsmath,amsfonts,amssymb,graphicx,epsfig,pdflscape}
\usepackage[usenames]{color}
\usepackage{pstricks}
\usepackage{arydshln,cite}

\numberwithin{equation}{section}

\definecolor{red}{rgb}{1,0,0}
\definecolor{lightblue}{rgb}{.61,.61,1}
\definecolor{midblue}{rgb}{.7,.7,1}
\definecolor{lightlightblue}{rgb}{.85,.85,1}
\definecolor{lightestblue}{rgb}{.96,.96,1}
\definecolor{lightpurple}{rgb}{1,.65,1}
\definecolor{pink}{rgb}{1,.65,.65}

\newcommand{\nc}{\newcommand}

\nc{\colend}{\normalcolor}

\newcommand{\ie}{i.e\mbox{.}\xspace}

\nc{\be}{\begin{equation}}
\nc{\ee}{\end{equation}}
\nc{\bea}{\begin{eqnarray}}
\nc{\eea}{\end{eqnarray}}
\def\vvdots{\mathinner{\mkern1mu\raise1pt\vbox{\kern7pt\hbox{.}}\mkern2mu
  \raise4pt\hbox{.}\mkern2mu\raise7pt\hbox{.}\mkern1mu}}
\nc{\rnc}{\renewcommand}
\rnc{\title}[1]{{\Large\bf\mbox{}\\\medskip#1\bigskip\medskip\\}}
\rnc{\author}[1]{{\large #1\smallskip\\}}
\nc{\address}[1]{{\em #1\medskip\\}}

\nc\SU{{\rm SU}}
\nc\su{{\rm su}}

\def\Tr{{\rm Tr\,}}
\nc\GD{\Delta}\nc\GS{\Sigma}
\nc\na{\rm{n.a.}}
\nc\cm{\checkmark}
\nc\sd{\rtimes}
\nc\Z{\Bbb{Z}}
\nc\C{\Bbb{C}}
\nc\one{{\bf{1}}}

\def\Nh{\hat N}\def\N{\hat N} 

\nc\Ga{\alpha}\nc\Gd{\delta}\nc\Gk{\kappa}
\nc\Gl{\lambda}\nc\Gm{\mu}\nc\Gn{\nu}\nc\Go{\omega}\nc\Gr{\rho}\nc{\Gs}{\sigma}
\def\CC{{\cal C}}\def\CE{{\cal E}}\def\CS{{\cal S}}\def\CK{{\cal K}}

\def\go{\frak{g}}
\def\goh{\hat{\frak{g}}}
\nc\irr{irreducible}
\nc\rep{representation}
\def\Th#1#2{{\noindent \bf Theorem #1: }{\sl #2}}
\def\ommit#1{{}}

\def\CP{{\red P}} \def\CPk{{\CP}_+^k}\def\CP{P}
\begin{document}

\begin{titlepage}
\vspace*{\fill}

\begin{center}
\title{ On sums of  tensor and fusion multiplicities}
\medskip
\author{Robert Coquereaux} 
\address{Centre de Physique Th\'eorique (CPT),\\ CNRS UMR 6207 \\
 Luminy, Marseille, France}
\medskip
\author{Jean-Bernard Zuber}
\address{
Laboratoire de Physique Th\'eorique et Hautes Energies, \\
CNRS UMR 7589 and Universit\'e Pierre et Marie Curie - Paris 6, \\
4 place Jussieu, 75252 Paris cedex 05, France }
\bigskip\medskip

\begin{abstract}
\noindent  {The total multiplicity in the decomposition into irreducibles of the tensor
product $\lambda\otimes \mu$ of two irreducible representations of a simple Lie algebra is invariant under
conjugation of one of them $\sum_\nu N_{\Gl\mu}^{\ \ \nu}=\sum_\nu N_{\bar\Gl\mu}^{\ \ \nu}$. 
This also applies to the fusion multiplicities of affine
algebras in conformal WZW theories. In that context, the statement is equivalent to a property 
of the modular $S$ matrix, viz $\GS(\kappa):=\sum_\Gl S_{\Gl\Gk}=0$ if $\Gk$ is a complex representation.
Curiously, this vanishing of $\GS(\kappa)$ also holds when $\kappa$ is a quaternionic representation.
We provide proofs of all these statements. These proofs  rely on a case-by-case analysis, 
maybe overlooking some hidden symmetry principle. We also give various illustrations of these 
properties in the contexts of boundary conformal field theories, integrable quantum field theories 
and topological field theories.}
\end{abstract}
\end{center}
\vspace*{20mm}
\vspace*{\fill}
\end{titlepage}

\section*{Introduction}

In the course of investigations of algebraic features of conformal
theories,
we have encountered a seemingly unfamiliar property of sums of
tensor product or fusion multiplicities of irreducible representations of
simple
or affine Lie algebras, and an associated property of the modular
$S$-matrix
in the affine algebra case. Let $N_{\Gl\Gm}^{\ \ \Gn}$ the multiplicity of
the irrep of weight
$\Gn$ in the tensor product of those of weights $\Gl$ and $\Gm$.
(Notations will be presented
with more care in the following section.) It is a commonplace to say  that
$N_{\Gl\Gm}^{\ \ \Gn}=N_{\bar\Gl\bar\Gm}^{\ \ \bar\Gn}$
 and that $N_{\Gl\Gm}^{\ \ \Gn}=N_{\bar\Gn\Gm}^{\ \ \bar\Gl}$,  
where $\bar\Gl$ is the complex conjugate weight of $\Gl$, and hence that $\sum_\Gn
N_{\Gl\Gm}^{\ \ \Gn}$
is invariant under the simultaneous conjugation of $\Gl$ and $\Gm$. We
claim that
the latter sum is also invariant under a single conjugation $\Gl\to \bar
\Gl$ : \quad
$\sum_\nu N_{\Gl\mu}^{\ \ \nu}=\sum_\nu N_{\bar\Gl\mu}^{\ \ \nu}$
(Theorem 1). The paper
consists of variations on that theme.

Here is the  layout of the paper.
The main results are presented in Sect 1 as a sequence of four theorems.
Theorem 1  is the above property for  tensor product multiplicities and Theorem 2
deals with the same property for fusion coefficients
within affine algebras. Theorem 3 and 4 assert that the sum
$\GS(\kappa):=\sum_\Gl S_{\Gl\Gk}$
vanishes  if $\Gk$ is a complex (Th 3) or quaternionic (Th 4)
representation.  Proofs of Theorems 1, 2, 3 and 4 will be given
in Sect 2, 3, 4 and 5, respectively. Sect 6 is a short discussion of
 cancellations of  $\GS(\kappa)$ that may also occur when  $\Gk$ is real, 
Sect 7 shows what may happen in finite groups, and
Sect 8 presents a few applications or illustrations of our properties in
various contexts, together with  some final comments. 
Appendices gather  lengthy details of our  proofs, and some useful tables.


\section{The main results}
Let $\go$ be a  finite dimensional  simple Lie algebra of rank $n$. 
Each of its  finite dimensional   \irr\ \rep s (irreps) is labelled by a highest weight (h.w.) $\Gl$. 
By a small abuse of notation, we refer to that \rep\ as \rep\ $\Gl$. Throughout this paper, we shall denote 
$[\Gl]$ the weight system of irrep $\Gl$.
Let $N_{\Gl\Gm}^{\ \ \Gn}$ denote the multiplicity of irrep $\Gn$ in the decomposition 
of the Kronecker product $\Gl\otimes \Gm$. Let $\bar \Gl$ denote the  \rep\ conjugate to $\Gl$.

\smallskip
\Th{1}{For a given pair $(\Gl,\Gm)$ 
of irreps of the simple Lie algebra $\go$,  the total multiplicity $\sum_\Gn N_{\Gl\Gm}^{\ \ \Gn}$ satisfies 
\be \label{summul}\sum_\Gn N_{\Gl\Gm}^{\ \ \Gn}=\sum_\Gn N_{\bar\Gl\Gm}^{\ \ \Gn}\ee
Equivalently, since $N_{\Gl\Gm}^{\ \ \Gn}=N_{\bar\Gn\Gm}^{\ \ \bar\Gl}$,
\be \label{equiv}
 \sum_{\Gl} N_{\Gl\mu}^{\ \ \nu}=\sum_\Gl N_{\Gl\mu}^{\ \ \bar\nu}
 \ee}
 
 \smallskip
\noindent Of course the Theorem is non-trivial only in cases where $\go$ has complex representations, 
{\it i.e.} $\go=A_n$, $D_{n=2s+1}$ or $E_6$.
Although this looks like a classroom exercise in group theory, we couldn't find either a reference 
in the literature or a simple  and compact argument and we had to resort to a case by case analysis, see Sect  \ref{sec:proofoftheorem1} below.
Note also that this property is not a trivial consequence of the general representation theory of groups; 
in particular, it does  not hold in general in finite groups, see Sect \ref{sec:finitesubgroupsection} below for counterexamples based on finite subgroups of $\SU(3)$.

\smallskip

Theorem 1 is also valid for the {\it fusion} multiplicities of integrable \rep s of affine Lie algebras taken at some level $k$. 
Such representations are the objects of a fusion category with a finite number of  simple objects that will just be called  {\it irreps}, for short.
These simple objects (and the category itself) could also be built in terms of irreducible representations of quantum groups at roots of unity that have non-vanishing quantum dimension.
One sometimes refers to this framework by saying that we consider the fusion category defined by $\go$ at level $k$, but for definiteness, when needed, we shall use the language of affine algebras, and 
denote $\goh_k$ the affine algebra of type $\go$ at the finite integer level $k$.
Then we have,  using the notation $\N_{\Gl\Gm}^{\ \ \Gn}$ for the fusion multiplicities, ( for completeness
a label $k$ should be appended to this notation but will be omitted)

\smallskip
\Th{2}{  Eq. (\ref{summul}) (or (\ref{equiv})) is valid for any pair $(\Gl,\Gm)$ of irreps of the fusion category defined by  $\goh_k$ at level $k$}
\be\label{summulaff}
\sum_\Gn \N_{\Gl\Gm}^{\ \ \Gn}=\sum_\Gn \N_{\bar\Gl\Gm}^{\ \ \Gn}  \ee
%
\be \label{equivaff}  \sum_{\Gl} \N_{\Gl\mu}^{\ \ \nu}=\sum_\Gl \N_{\Gl\mu}^{\ \ \bar\nu}  \,. \ee

 \smallskip 
 Part of the proof given in Sect \ref{sec:proofoftheorem1}  can be used in that case, but the discussion needs nevertheless to be extended, so that the proof of Theorem 2 is given in Sect  \ref{sec:proofoftheorem1bis}. Notice that the theorem for simple algebras follows from that  for affine algebras, provided the level is chosen large enough. 

Now in that context of affine algebras, the multiplicities $\N_{\Gl\Gm}^{\ \ \Gn}$ are given by 
the Verlinde formula \cite{Ve} in terms of the unitary modular $S$ matrix
\be \label{verlinde}
\N_{\Gl\Gm}^{\ \ \Gn}=\sum_\Gk {S_{\Gl\Gk}S_{\Gm\Gk}S^*_{\Gn\Gk}\over S_{0\Gk}}\ ,\ee
where the weight $0$ refers to the identity \rep. 
We recall that the matrix $S$ is symmetric $S_{\Gl\Gk}=S_{\Gk\Gl}$ and 
satisfies the following properties
\be S^\dagger=S^{-1}=S^3=S C=C S \ee
where $C=S^2$ is the conjugation matrix: $C_{\Gl\Gl'}=\Gd_{\Gl'\bar\Gl}$ from which it follows that
\be
\label{conj}
S_{\bar\Gl\Gk}=S_{\Gl\bar\Gk}=S_{\Gl\Gk}^*\ .
\ee
Then we have the (apparently) stronger constraint on $\GS(\Gk):=\sum_\Gn S_{\Gn\Gk}$

\bigskip
\Th{3}{ $\ { \GS(\Gk):=}\sum_\Gn S_{\Gn\Gk}= 0 $ if $\Gk\ne \bar \Gk$\ .}

\smallskip
That Theorem 3 implies Theorem 2 is readily seen:
\bea\sum_\Gn \N_{\Gl\Gm}^{\ \ \Gn}&=&\sum_{\Gk}{S_{\Gl\Gk}S_{\Gm\Gk}\sum_\Gn S^*_{\Gn\Gk}\over S_{0\Gk}}
\ \buildrel {\rm (Th. 3)}\over{=}\ \sum_{\Gk=\bar \Gk}{S_{\Gl\Gk}S_{\Gm\Gk}\sum_\Gn S^*_{\Gn\Gk}\over S_{0\Gk}}\nonumber \\
&\buildrel (\ref{conj})\over{=}&\sum_\Gn \sum_{\Gk=\bar \Gk}{S_{\bar \Gl\Gk}S_{\Gm\Gk}S^*_{\Gn\Gk}\over S_{0\Gk}}
=\sum_\Gn \N_{\bar\Gl\Gm}^{\ \ \Gn}\ . \label{imply}
\eea
As we shall see below, (Sect 4),  Theorem 3 also follows from Theorem 2, so that the two statements are in fact
equivalent.

Theorem 3 states that $\sum_\Gn S_{\Gn\Gk}$ vanishes if $\Gk$ is a complex \rep. 
Or  equivalently it  may be non-zero 
only if $\Gk$ is  self-conjugate. As is well known
this covers two cases, {\it real} \rep s and {\it quaternionic}, also known as  {\it pseudoreal}, \rep s.
We show in Sect \ref{sec:quaternionic} that 
 
 \bigskip
\Th{4}{\sl Let $\Gk$ be an irrep of $\goh_k$. If  $\Gk$ is of quaternionic type, the sum $\GS(\Gk)=
\sum_\nu S_{\nu \Gk}$ vanishes.}\\

The sum $\sum_\Gn S_{\Gn\Gk}$ may thus be non-zero only if $\Gk$ is a real representation. Actually this sum can sometimes vanish, even for real representations, either because it is forced by some automorphism of the Weyl alcove, or because of some accidental property of the representation $\Gk$. We return to this question in Sect \ref{sec:realrepresentations}.


\section{Sum of multiplicities (classical case). Proof of Theorem 1}
\label{sec:proofoftheorem1} 

The proof will be done in two steps. We first prove it for $\Gl$ one of the fundamental \rep s $\Go_p$, $p=1,\cdots n$, 
and $\Gm$ arbitrary; and then use the fact that any $N_{\Gl}$ is a polynomial in $N_{\Go_1},\cdots, N_{\Go_n} $.

\smallskip
\noindent {\bf Lemma 1: }{\sl Theorem 1 holds for  any fundamental weight $\Gl=\Go_p$.} 

\noindent We recall a well known method  of calculation of the multiplicities $N_{\Gl\Gm}^{\ \ \Gn}$ for two given
h.w. $\Gl$ and $\Gm$, often called the Racah--Speiser algorithm \cite{RaSp, Fu, DFMS}. 
Here and below we write the components  of weights along the basis of fundamental weights (Dynkin labels).
Let $\Gr$ stand for the Weyl vector, \ie the sum of all fundamental weights (or half the sum of positive roots) of $\go$.
Consider the set 
of weights $\Gs=\Gl'+\Gm + \Gr$ where $\Gl'$ runs over the weight system $[\Gl]$ of the irrep of h.w. $\Gl$. 
Three cases may occur:
\begin{itemize}
\item{i)} if all Dynkin labels of $\Gs$ are positive,
$\Gl' + \Gm$ contributes to the sum over h.w. $\nu$ with a multiplicity 
equal to the multiplicity of $\sigma$ (\ie of $\lambda'$); 
\item{ii)} 
if $\Gs$ or any of its images under the Weyl group has a vanishing Dynkin label, 
i.e. if $\Gs$ is on the edge of a Weyl chamber,
$\Gl' + \Gm$  does not contribute to the sum over $\nu$;
\item{iii)} if $\Gs$ 
has negative  (but no vanishing) Dynkin labels,  and is not of the type discussed in case (ii), it may be mapped inside the fundamental Weyl chamber  by  a unique  element $w$ of the Weyl group. The weight $w[\Gs] - \rho$ contributes with a multiplicity sign$(w)$ to the sum over $\nu$. 
\end{itemize}

This is summarized in the formula
\be\label{rasp}
N_{\Gl\Gm}^{\ \ \Gn}=\sum_{\Gl'\in [\Gl]}\sum_{w\in W \atop  w[\Gl'+\Gm + \Gr]-\Gr\in \CP_+} {\rm sign}(w)\,
\Gd_{\nu, w[\Gl'+\Gm + \Gr]-\Gr}
\ee
in which  $\CP_+$ is the fundamental Weyl chamber ($\Gn\in \CP_+ \Leftrightarrow \nu_i \ge 0\ \forall  i=1,\cdots n$).

Remarks. {1.} In practice, it may not always be immediately obvious to discover that a shifted weight $\sigma$ belongs to the edge of a Weyl chamber and therefore trivially contributes to the problem,  but one can easily discard  at least those $\sigma$ with one or several Dynkin labels equal to $0$ since they obviously belong to the walls of the fundamental chamber. In any case the trivial $\sigma$'s that would not be recognized as such will be mapped, at a later stage, to the walls of the fundamental Weyl chamber, and they can be removed then. Note that in formula (\ref{rasp}) these cases of type (ii) automatically cancel out, as they contribute with two Weyl
elements of opposite signatures.

{2.} The irreps $\nu$ that appear in the decomposition into irreps of the tensor product $\lambda \otimes \mu$ are obtained (together with their multiplicities) from the non-trivial contributions  (i) and (iii). 
The same weight $\nu$ can sometimes be obtained both from (i) and (iii), possibly with different signs.  Its final multiplicity is the {\sl algebraic} sum of its partial multiplicities. 

{3.} Notice that, as a 
consequence of the above method,  the sum over $\nu$ of multiplicities $N_{\lambda \mu}^{\ \ \nu}$ should 
be smaller than the dimensions of any of the two irreps $\lambda$ and $\mu$  entering the tensor product,
 $\sum N_{\Gl\Gm}^{\ \ \Gn}\le \inf(\dim(\Gl), \dim(\Gm))$.\\

We shall now see that for all the complex fundamental \rep s of the $A$, $D$ and $E_6$ algebras (with one exception 
in $E_6$, see below), we are  in case (i) or (ii),
and that for $\Gl=\Go_p$ or $\Gl=\bar\Go_p$, the occurrences of (ii) are equinumerous, thus proving the Lemma.

For each of the fundamental \rep s $\Go_p$ of the  $A_n$ algebra ($p=1,\cdots,n$), for the spinorial \rep s\footnote{those are the only fundamental complex \rep s of the $D_{2s+1}$ case.}
$\Go_{n-1}$ and $\Go_n$  of the $D_n$ ($n=2s+1$)  algebra, and for the 
27-dimensional fundamental \rep s $\Go_1$ and $\Go_5$ 
of  $E_6$, the Dynkin labels of the weights $\Gl'$ of the weight system of $\Gl=\Go_p$ 
take the value $0$ or $\pm 1$. Thus after addition of $\Gr$ (whose Dynkin labels are all equal to 1),
the Dynkin labels of $\sigma=\Gl'+\Gr+\mu$ are never negative and the case (iii) above never occurs.
On the other hand, case (ii) occurs whenever some Dynkin label of $\mu$ vanishes while the corresponding
one in $\Gl'$ equals $-1$. It is easy to check by inspection that there  is the same number of
weights with $-1$ entries at given locations $1\le i_1<i_2<\cdots i_q\le n$ 
in the weight systems of any $\omega_p$ and $\bar\omega_p$.
For a given $\mu$, there is thus an equal number of occurrences of cases of type (ii) for the 
fundamental weights $\Go_p$ and $\bar\Go_p$. 

To complete the proof of Lemma 1, we still have to consider the case of 
the complex, 351-dimensional, \rep s $\Go_2$ and $\bar\Go_2=\Go_4$ of $E_6$ (notice that $\Go_2$ is also the antisymmetric tensor square of $\Go_1$).  This requires 
a particular analysis because the weight system of $\Go_2$ (or of $\Go_4$) contains weights with Dynkin labels equal to $-2$, so that when the corresponding label of $\mu$ vanishes, we are in the situation (iii). For the sake of clarity, this detailed 
discussion is relegated to Appendix \ref{appA1}.

\def\({\left(}
\def\){\right)}
\def\nn{\nonumber}

\medskip
\noindent {\bf Lemma 2: }{\sl Theorem 1 holds for any product of the fundamental representations.}

\noindent  In the following it will be convenient to use an alternative notation for the
multiplicities $N_{\Gl\Gm}^{\ \ \Gn}$ and to regard them as the $(\Gm,\Gn)$  entry of the matrix $N_\Gl$.
We have proved in  Lemma 1 that for any $p$
$$ \sum_{\nu}\left(N_{\Go_p}\right)_\mu^{\  \nu}=\sum_{\nu}\(N_{\bar \Go_p}\)_\mu^{\  \nu}\ .$$
This, together with the commutativity of the $N$ matrices, 
 entails that for any monomial $N_{\Go_{j_1}}\cdots N_{\Go_{j_q}}$
\bea \sum_{\nu}\(N_{\Go_{j_1}}\cdots N_{\Go_{j_q}}\)_\mu^{\  \nu} 
&=&\sum_{\nu'}\(N_{\Go_{j_1}}\cdots N_{\Go_{j_{q-1}}}\)_\mu^{\  \nu'}\sum_{\nu }\( N_{\Go_{j_q}}\)_{\nu'}^{\ \nu}\nn \\
&\buildrel {\rm (Lemma\ 1)}\over =&\sum_{\nu'}\(N_{\Go_{j_1}}\cdots N_{\Go_{j_{q-1}}}\)_\mu^{\  \nu'}\sum_{\nu }\( N_{\bar\Go_{j_q}}\)_{\nu'}^{\ \nu}\nn \\
&=&\sum_{\nu}\(N_{\bar\Go_{j_q}}N_{\Go_{j_1}}\cdots N_{\Go_{j_{q-1}}}\)_\mu^{\  \nu}=\cdots\nn \\
&=&\sum_{\nu}\(N_{\bar\Go_{j_1}}\cdots N_{\bar\Go_{j_q}}\)_\mu^{\  \nu}
\eea
which exhibits the product of the conjugate fundamental representations. 

2) Now
it is also well known \cite{Weyl, DFMS} that any irreducible representation may be obtained from a suitable combination
of tensor products of the fundamentals. Or in other words, any matrix $N_\Gl$ is some polynomial (with 
integer coefficients)\footnote{a generalized Chebyshev polynomial \cite{DFMS}.} of the commuting $N_{\Go_1},\cdots, N_{\Go_{n}}$: 
$N_\Gl=P_{\Gl}(N_{\Go_1},\cdots,N_{\Go_{n}})$
and  $N_{\bar\Gl}=P_{\Gl}(N_{\bar\Go_{1}},\cdots,N_{\bar\Go_{n}})$. 
Thus the property proved above for any monomial establishes the general statement and completes
the proof. 


\section{Sum of multiplicities (affine/quantum case). \\ Proof of Theorem 2}
 \label{sec:proofoftheorem1bis}

\def\rung{{rung}}\def\rung{level}

\subsection{Levels and automorphisms}
\label{sec:levelsandautomorphisms}
 \def\ssigma{\varkappa} 
 \def\ssigma{\zeta} 
 Let $\CPk$ be the set of integrable weights of the affine algebra $\goh$ at level $k$\cite{Ka}. Each 
 weight of $\CPk$ is completely specified by a dominant 
 weight $\Gl$ of the   underlying classical algebra $\go$,  restricted  by the 
condition $\CK(\Gl)\le k$ where $\CK$ is the linear form $\CK(\Gl):= \langle \Gl,\theta\rangle $ and  $\theta$ is the highest root of $\go$.
We shall call {\sl \rung}  of a weight $\Gl$  the integer  $\CK(\Gl)$. Therefore a weight  exists
in a representation of  level $k$ when its \rung\ is smaller than or equal to $k$.
 By another  slight abuse of notation, $\Gl$ will 
 denote both the weight of  $\goh$  and the corresponding weight in $\go$. 
  We refer to the subset of $\Gl$ such that 
 $\CK(\Gl) = k $ as ``the back wall" (of the Weyl alcove $\CPk$).  It is also convenient to introduce the additional
 Dynkin label $\Gl_0=k-\CK(\Gl) $ of the affine weight $\Gl$ : clearly $\Gl_0$ vanishes on the back wall. 
  
Each of the algebras $\goh$ with complex representations, i.e. $\hat A_n$, $\hat D_{2s+1}$ and $\hat E_6$, has the
well known properties
\begin{itemize} 
\item{} the set $\CPk$ of integrable weights at level $k$  is invariant under the action of an automorphism~$\ssigma$;
\item{} there exists a $\Z_N$-grading $\tau$  on the weights of $\CPk$ :  
$N=n+1$ for $\hat A_n$,  
$N=4$ for $\hat D_{2s+1}$ and 
$N=3$ for $\hat E_6$;   \ommit{\blue the weight $0$ has $\tau(0)=0$;}
\item{} the modular $S$-matrix satisfies the relation \cite{Be} 
\be \label{expophase}
S_{\ssigma(\Gm)\Gk}=e^{2\pi i \tau(\Gk)/N} S_{\Gm\Gk}\,.
\ee
 \end{itemize}
The value of the 
\rung\ $\CK(\Gl)$ may be calculated easily from the expansion of the highest root  $\theta$ in terms of 
simple roots (Coxeter--Kac labels): $\theta=(1,1,\cdots,1),\ (1,2,2,\ldots,2,1,1),\ (1,2,3,2,1,2)$ for 
$A_n, D_{2s+1},\ E_6$ respectively.
The expressions  of $\CK(\Gl)$, the automorphisms, the $\Z_N$-grading and the conjugates in 
the three above algebras are gathered in Appendix \ref{sec:automorphisms}.
One can check on these expressions that the \rung\ of a weight is invariant by conjugation: 
$\CK(\Gm)=\CK(\bar\Gm)$. 
Moreover the automorphism $\ssigma$ and the complex conjugation satisfy the consistency relation 
\be
\ssigma(\bar \Gm)=\overline{\ssigma^{-1}(\Gm)}\,,
\label{auto-conj}
\ee
and by iteration
\be \ssigma^p(\bar \Gm)=\overline{\ssigma^{-p}(\Gm)}\qquad \forall p\,.
\label{auto-conj-iter}
\ee
For the $A_n$ algebra one finds that
$\CK(\ssigma(\Gm))= k-\Gm_n $, $\CK(\ssigma^{-1}(\Gm))= k-\Gm_1 $
and more generally 
 \be\label{level-sigma} 
 \CK( \ssigma^{-p}(\Gm))= k-\Gm_p\,, \ee 
 while for the $D_{2s+1}$ case,
 \be \label{lev-sigmaD}
 \CK( \ssigma(\Gm))= k-\Gm_{2s}\,,\quad  \CK(\ssigma^{\pm 2}(\Gm))= k-\Gm_1\,,  \quad  \CK( \ssigma^{-1}(\Gm))
 = k-\Gm_{2s+1}\,,\ee
 and for the $E_6$ case
 \be  \CK(\ssigma(\Gm))= k-\Gm_5\,,\quad  \langle \CK(\ssigma^{-1}(\Gm))= k-\Gm_1\,.\ee
\medskip
$\ssigma$ is an automorphism of the fusion rules as a consequence of (\ref{verlinde}) and (\ref{expophase})
\be \N_{\Gl \ssigma(\Gm)}^{\ \ \ \ \ssigma(\nu)}=\sum_\Gk {S_{\Gl\Gk} S_{\ssigma(\Gm)\Gk} S_{\ssigma(\nu)\Gk}^*\over S_{0\Gk}}=
\sum_\Gk {S_{\Gl\Gk} S_{\Gm\Gk} S_{\nu\Gk}^*\over S_{0\Gk}}=\N_{\Gl \Gm}^{\ \ \nu} \,. \ee
This implies that the sum of multiplicities satisfy 
\be \sum_\Gn \N_{\Gl \Gm}^{\ \ \nu}= \sum_\Gn \N_{\Gl \ssigma(\Gm)}^{\ \ \ \ \nu}\,.  \label{sigmasum}\ee
\ommit{\blue  Also, applying the automorphism $\zeta$ on the summation variable $\kappa$ of Verlinde's formula yields \be\label{selrule} \N_{\Gl \Gm}^{\ \ \nu} \ne 0 \hbox{only if } \tau(\Gl)+\tau(\Gm) =\tau(\Gn)\,\mod N\,. \ee}

\subsection{Fusion coefficients}

 There are several alternative routes to determine the fusion coefficients. Let us quote three of them.
The first  is the Verlinde formula (\ref{verlinde}), which relies on the knowledge of the modular $S$-matrix.

\def\hW{\widehat{W}}
Secondly one may use  an affine generalization of the Racah--Speiser algorithm described in 
eq. (\ref{rasp}) 
\be\label{raspaff}
\Nh_{\Gl\Gm}^{\ \ \Gn}=\sum_{\Gl'\in [\Gl]}\sum_{w\in \hW \atop  w[\Gl'+\Gm + \Gr]-\Gr\in \CPk} {\rm sign}(w)\,
\Gd_{\nu, w[\Gl'+\Gm + \Gr]-\Gr}
\ee  
The  modification is twofold : the fundamental Weyl chamber $\CP_+$ is replaced by $\CPk$, the Weyl 
alcove $\CPk$  of level $k$; and  the sum runs now over elements of the {\it affine} Weyl group $\hW$,
of which the reflection $s_0$ across the shifted back wall 
is the new generator. 
What is referred to as the shifted back wall is the hyperplane
of equation
 $\CK(\Gl)= k+ h^\vee$, and the reflection $s_0$ acts according to 
$s_0[\Gl]=\Gl+ (k+ h^\vee-\CK(\Gl)){2\theta\over \langle \theta,\theta\rangle}$, 
where $h^\vee=1+\langle \Gr,\theta\rangle$ is the dual Coxeter number.
Just like in Sect 2,  weights $\Gl'$ which are such that $\Gl'+\Gm + \Gr$ 
lies either on an ordinary wall of the Weyl chamber, or on the shifted
back wall, or  on one of their images by $\hW$, do not contribute to the sum.

Thirdly, the fusion coefficients  $\Nh_{\Gl\Gm}^{\ \ \nu}$ and  the
ordinary multiplicities $N_{\Gl\Gm}^{\ \ \nu}$ occurring in the  ``horizontal'' algebra $\go$
are related by the Kac--Walton formula \cite{Kac-Wa}, 
\be \label{kac-walton} \forall\, \Gl,\Gm,\Gn\in \CPk\qquad
\Nh_{\Gl\Gm}^{\ \ \nu} =\sum_{w\in \hW\atop 
w[\nu+\rho]-\rho\,\in  {\CP}_+} {\rm sign}(w)\,  
N_{\Gl\Gm}^{\ \ w[\nu+\rho]-\rho}\,. \ee

 As far as the proof of Theorem 2 is concerned, the first method (Verlinde formula)
 does not seem appropriate, unless some additional properties of that matrix (in fact our Theorem 3) are proved beforehand.  
 On the other hand, repeating the method of Sect 2 with the affine version of the Racah--Speiser algorithm 
 leads in a straightforward  way to a proof, as we shall see in the next subsection.
Using the results of Sect 2 on sums of tensor product multiplicities together with (\ref{kac-walton}) 
 and the automorphism $\ssigma$ of section \ref{sec:levelsandautomorphisms}
 is another tantalizing possibility, which however seems to be applicable only to a subset of cases. We return
 to this point at the end of next subsection.

\subsection{Proof of Theorem 2}

As in Sect 2, we take $\Gl$ to be the highest weight of one of the complex fundamentals of the affine algebra $\goh$ with $\go=A_n,\  
D_{n=2s+1}$  or $E_6$. 
Again, in the latter case, we treat separately the weights  $\Go_2$ and $\Go_4$  (their level is $2$).
Each of   the other cases ($\Gl=\Go_p$, $p=1,\cdots, n$, in $A_n$,    
$\Go_{2s}$ or $\Go_{2s+1}$ in $D_{n=2s+1}$,  and $\Go_1$ or $\Go_5$ in $E_6$) has a   \rung\ $\CK(\Go_p)=1$,
  and all the weights $\Gl'$ of the representation $\Gl$ have a \rung\ $\CK(\Gl')=\pm 1$ or $0$, as is readily 
 checked on their expression.
 \ommit{{ {\small \green Je d\'etaille: pour $A_n$, 
 poids g\'en\'erique de la repr\'esentation $\omega_p$ : $e_{i_1i_2\cdots i_p}=e_{i_1}+e_{i_2}+\cdots+e_{ i_p}$
 avec $1\le i_1<i_2<\cdots i_p\le n+1$ et $e_i=\Go_i-\Go_{i-1}$ avec $\Go_0:=\Go_{n+1}:=0$.
 Donc $\CK(e_1)=1$, $\CK(e_i)=0$ pour $i=2,\cdots, n$, $\CK(e_{n+1})=-1$. Par lin\'earit\'e, 
 $\CK(e_{i_1i_2\cdots i_p})=\pm 1$ or $0$.
Par exemple, pour le hw $\Go_p=e_{12\cdots p}$, $p\le n$, $\CK(\Go_p)=1$.
Pour $E_6$, par inspection : les 35 poids de $\Go_5$ ont $\CK=0,\pm 1$. Et pour $D_{2s+1}$, 
\dots c'est aussi vrai !\dots OK ?}}}

We then follow the same steps as in Sect 2: for any weight 
$\Gm\in \CPk$, hence with all its Dynkin labels (including the affine label $\Gm_0$) non-negative, 
and for any $\Gl'\in[\Gl=\Go_p]$,  one sees that $\Gs=\Gl'+\Gm+\Gr$ has non-negative 
 Dynkin labels $\Gs_i$, $i=1,\cdots, n$ and  likewise
\be \label{sigma0}\Gs_0=k+h^\vee-\CK(\Gs)=(k-\CK(\mu))+(1-\CK(\Gl'))\ge 0 \,.\ee 
Hence no non-trivial $w$ has to be applied to 
$\Gs$ to bring it back (after subtraction of $\Gr$) to $\CPk$. But some of these $\Gs$ may lie on a wall and will not
contribute to the sum in (\ref{raspaff}),  and this 
occurs whenever one or several of the Dynkin labels $\Gm_i$, $i=0,\cdots, n$ vanish. 
In view of the discussion of Sect 2 for the finite case, it suffices to examine the situation
 when $\Gs$ lies on the shifted back wall, i.e.   $\Gs_0$ vanishes, and (\ref{sigma0}) says
  this occurs  whenever $\Gm$ lies on the back wall of $\CPk$ {\it and\/}  $\CK(\Gl')=+1$.
  Since for any $\Gl'$ of level 1, its conjugate $\bar \Gl'$ has also level 1, 
the number of these occurrences is the same for $\Gl=\Go_p$ and $\bar \Go_p$, 
and  like  in the finite case of Sect 2, this implies the equality
$\sum_\Gn \Nh_{\Go_p\Gm}^{\ \ \ \Gn}=\sum_\Gn \Nh_{\bar\Go_p\Gm}^{\ \ \ \Gn}$.
 The case of $\Gl=\Go_2$ or $=\Go_4$ for $E_6$ has again to be treated separately  and  will be 
relegated to Appendix \ref{appA2}.

Once it has been established for $\Gl$ one of the fundamentals, 
 Theorem 2 then follows  in general from the fact that the fusion ring is polynomially generated by the fundamental fusion matrices $\Nh_{\Go_p}$\cite{DFMS}.
 \bigskip

 An alternative route using the Kac--Walton formula (\ref{kac-walton})
is also applicable to the $A_n$ case (and also to $D_{2s+1}$ case {\it at odd level $k$}).
The method stems from the observation that when $\Gl$ or $\Gm$ are sufficiently off the back wall,  so that all $\Gn$ such that $N_{\Gl\Gm}^{\ \ \nu}\ne 0$
 are themselves in $\CP_+^k$, only $w=1$  contributes to the sum in (\ref{kac-walton}) and $ \Nh_{\Gl\Gm}^{\ \ \nu}$  does not differ from $ N_{\Gl\Gm}^{\ \ \nu}$.
Unfortunately the method does not seem  to be of general validity and we have thus to rely on the more systematic proof given previously.

 
\ommit{
{\sl An alternative route using the Kac--Walton formula (\ref{kac-walton})}\\
 The formula (\ref{kac-walton}) implies
   that when $\Gl$ or $\Gm$ are sufficiently off the back wall,  so that all $\Gn$ such that $N_{\Gl\Gm}^{\ \ \nu}\ne 0$
 are themselves in $\CP_+^k$, only $w=1$  
 contributes to the sum in (\ref{kac-walton}) and  
 $ \Nh_{\Gl\Gm}^{\ \ \nu}$    does not differ from $ N_{\Gl\Gm}^{\ \ \nu}$.\\
 Let us first restrict ourselves to the $A_n$ case and to its fundamental 
 \rep s $\Go_p$.
If  $\Gm$ is not on the back wall, $\CK(\Gm)<k$, and since $\CK(\Gl=\Go_p)=1$, 
$\CK(\Gl+\Gm)\le k$,   the only term contributing in (\ref{kac-walton})
is $w=1$  and  $\Nh_{\Gl\Gm}^{\ \ \nu}= N_{\Gl\Gm}^{\ \ \nu}$. 
  The same holds for the conjugate $\bar \mu$,  hence (\ref{summulaff}) follows from (\ref{summul}). 
The ``dangerous cases" occur when $\Gm$  is on  the back wall, {\it i.e.} $\CK( \Gm) =k $. .
But if this is the case, we may use $\ssigma$ 
to rotate $\Gm$ away from the back wall. If $\Gm_p$ is the first non-vanishing Dynkin label of $\Gm$, 
(\ref{level-sigma}) tells us that $\ssigma^{-p}(\Gm)$ is off the back wall, then $\N_{\Gl \ssigma^{-p}(\Gm)}^{\qquad \nu}=N_{\Gl \ssigma^{-p}(\Gm)}^{\qquad \nu}$  by the previous discussion and we may write
\bea \sum_\Gn \N_{\Gl \Gm}^{\ \ \nu} & \buildrel (\ref{sigmasum})\over{=}& \sum_\Gn \N_{\Gl \ssigma^{-p}(\Gm)}^{\qquad \ \ \nu}= \sum_\Gn N_{\Gl \ssigma^{-p}(\Gm)}^{\qquad \ \ \nu} \\
&\buildrel (\ref{summul})\over{=}& \sum_\Gn N_{\Gl \overline{\ssigma^{-p}(\Gm)}}^{\qquad \quad \nu}\buildrel (\ref{auto-conj})\over{=}\sum_\Gn N_{\Gl \ssigma(\bar\Gm)}^{\qquad \nu}= \sum_\Gn  \N_{\Gl \ssigma(\bar\Gm)}^{\quad\ \ \nu}\buildrel (\ref{sigmasum})\over{=}\sum_\Gn \N_{\Gl \bar\Gm}^{\ \ \nu}\,, \nonumber\eea
which completes the proof of (\ref{summulaff}) for $\Gl$ one of the fundamentals in the $A_n$ case.
 The same argument applies to the $D_{2s+1}$ case {\it at odd level $k$}:
 its two complex fundamental representations  have highest weights
$\Go_{2s}$ and $\Go_{2s+1}$ of \rung\   1; if $\Gm$ lies on the back wall, $\CK(\Gm)=
\Gm_1+2\sum_{i=2}^{2s-1}\Gm_i+\Gm_{2s}+\Gm_{2s+1}= k$, $\Gm_1,\Gm_{2s}$ and $\Gm_{2s+1}$ cannot
be all vanishing (since $k$ is assumed odd), hence according to (\ref{lev-sigmaD}),
either  $ \ssigma(\Gm)$, or $ \ssigma^{\pm 2}(\Gm)$ or $ \ssigma^{-1}(\Gm)$ is off the back wall. But the 
argument fails for 
$D_{2s+1}$  at even level if $\Gm_1=\Gm_{2s}=\Gm_{2s+1}=0$  and likewise for  $E_6$ if 
$\mu_1=\mu_5=0$.
We conclude that this tentative  alternative route making use of the automorphism
$\ssigma$ is of limited application and we have to rely to the more systematic proof given above.}

\section{Proof of Theorem 3}
 \label{sec:proofoftheorem2}

\noindent
We want to show (Th 3) that  if $\Gk\ne \bar \Gk$, then  $\Sigma(\Gk)=\sum_\Gn S_{\Gn\Gk}= 0 $.

\medskip
\noindent
If $\Gk\ne \bar \Gk$, there are two cases,  either $\tau(\Gk)$ vanishes, or it does not.
The proof splits then naturally into two parts.

First observe that for any $\Gk$ of non-vanishing $\tau$, $\sum_\Gl S_{\Gl\Gk}=0$.
Indeed 
\be \label{expophasesum}
 \sum_\Gl S_{\Gl\Gk}= \sum_\Gl S_{\ssigma(\Gl)\Gk} =e^{2\pi i \tau(\Gk)/N} \,  \sum_\Gl S_{\Gl\Gk} \,.
\ee

As we shall now see,  if $\Gk$ is such that $\sum_\Gl S_{\Gl\Gk}\ne 0$, then for any $\mu$, we have
$S_{\mu\Gk} = S_{\mu\bar\Gk}\,$, and for $\Gk \neq \overline{\Gk}$, this leads to a contradiction.
Therefore, if $\Gk$ is such that $\sum_\Gl S_{\Gl\Gk}\ne 0$, then $\Gk = \overline{\Gk}$. Equivalently, if $\Gk \neq \overline{\Gk}$, then $\sum_\Gl S_{\Gl\Gk} =  0$, even if $\tau(\Gk)$ vanishes.

\noindent
Completing the proof therefore requires two small lemmas that we now discuss in detail.

\bigskip

Verlinde formula (\ref{verlinde}) implies
\be
S_{\Gl\Gk}S_{\mu\Gk}=\sum_{\nu}\N_{\Gl\mu}^{\ \ \nu} S_{\nu\Gk}S_{0\Gk}\,.
\ee
and we have  proved that $\sum_{\Gl} \N_{\Gl\mu}^{\ \ \nu}=\sum_\Gl \N_{\Gl\mu}^{\ \ \bar\nu}$, see  (\ref{equivaff}).
Therefore, for any $\Gk$,
\bea\hskip-8mm
(\sum_\Gl S_{\Gl\Gk})S_{\mu\Gk}&=&\sum_{\nu}(\sum_\Gl \N_{\Gl\mu}^{\ \ \nu}) S_{\nu\Gk}S_{0\Gk} 
=\sum_{\nu}(\sum_\Gl \N_{\Gl\mu}^{\ \ \bar\nu}) S_{\nu\Gk}S_{0\Gk}\nonumber\\
&=&\sum_{\nu}(\sum_\Gl \N_{\Gl\mu}^{\ \ \nu}) S_{\bar\nu\Gk}S_{0\Gk}
=\sum_{\nu}(\sum_\Gl \N_{\Gl\mu}^{\ \ \nu}) S_{\nu\bar\Gk}S_{0\Gk}\\
&=&\sum_{\nu}(\sum_\Gl \N_{\Gl\mu}^{\ \ \nu}) S_{\nu\bar\Gk}S_{0\bar\Gk}
=\sum_\Gl S_{\Gl\bar\Gk}S_{\mu\bar\Gk}=\sum_\Gl S_{\bar\Gl\Gk}S_{\mu\bar\Gk}=(\sum_\Gl S_{\Gl\Gk})S_{\mu\bar\Gk}\nonumber
\eea
where we used the fact that  $S_{0\bar\Gk}=S_{0\Gk}$ is real (it is a quantum dimension up to a real factor $S_{00}$),
 and that summations over $\nu$ or $\bar\nu$ are equivalent. Therefore we have proved

\noindent {\bf Lemma 3: }{\sl  For any $\Gk$ 
such that $\sum_\Gl S_{\Gl\Gk}\ne 0$ (hence of vanishing $\tau$), and for any $\mu$, we have
\be\label{truc}
S_{\mu\Gk} = S_{\mu\bar\Gk}\,.
\ee}
 To complete the proof, we have to show that  this situation cannot occur for $\Gk$ complex.

\noindent {\bf Lemma 4: }{\sl For any  
complex $\Gk$, {\it i.e.} $\Gk\ne \bar\Gk$,  
 there exists a weight $\mu\in \CPk$ such that 
\be S_{\mu\Gk}\ne (S_{\mu\Gk})^*=  S_{\mu\bar\Gk} \,.
\ee}
Note this holds irrespectively of whether $\tau(\Gk)$ vanishes or not. \\
Proof. For such a $\Gk\ne \bar\Gk$, (the h.w.  of a complex representation), 
the fusion matrices $\N_\Gk$ and $\N_{\bar\Gk}$ are different, since  $(\N_\Gk)_0^{\ \Gk}=1$ 
\ommit{\blue (by (\ref{selrule}))}
whereas
$(\N_{\bar\Gk})_0^{\ \Gk}=0$. But these two matrices are diagonalized in the same basis
through Verlinde's formula, with eigenvalues $S_{\Gk\mu}/S_{0\mu}$, resp. $S_{\bar\Gk\mu}/S_{0\mu}$.
Thus there is at least one distinct pair of eigenvalues $S_{\Gk\mu}\ne S_{\bar\Gk\mu}$. The lemma
is proved.

\bigskip 
Lemma 4, together with Lemma 3 (\ref{truc}), implies that $\sum_\Gl S_{\Gl\Gk}\ne 0$ is only possible if $\Gk=\bar\Gk$, and this
completes the proof of Theorem 3.


\medskip
\noindent {\bf  Comment}\\
The previous discussion was needed to handle the general case where the representation $\kappa$ is complex, but let us remember that for those particular complex representations of non-vanishing $\tau$, the proof of the vanishing of   $\sum_\lambda S_{\lambda \Gk}$ is immediate.
In the case of $A_n$, such a simplified proof can be  given 
for instance if $\kappa$ is a fundamental representation, and more generally when $\sum_j  j \, \kappa_j  \neq 0 \mod n+1$.
In the case of  $E_6$, assuming $\kappa$ complex, ie   $\kappa_1 \neq \kappa_5$ or  $\kappa_2 \neq \kappa_4$,  such a simplified proof can also be  given 
for the complex fundamentals $(100000)$, $(010000)$ and their conjugates $(000010)$, $(000100)$, and more generally when $2 \kappa_1+\kappa_2+2\kappa_4+\kappa_5 = 1, 2 \mod 3$.


\section{The case of quaternionic \rep s} 
\label{sec:quaternionic}

\subsection{The case of $\su(2)$}

For the $\widehat{su}(2)_k$ algebra, the integrable weights are $\Gl\in \{0,1,\cdots, k\}$. Denote $h=k+2$ for brievity.
Then $S_{\Gl\Gk}=\sqrt{2\over h} \sin{(\Gl+1)(\Gk+1)\pi\over h}$ and
$$ \sqrt{h\over 2}\sum_{\Gl=0}^{k} S_{\Gl\Gk}=-{\cos{\pi (\Gk+1)(2h-1)\over 2 h}-\cos{\pi (\Gk+1)\over 2h}\over 2 \sin{\pi (\Gk+1)\over 2h}}
={(1-(-1)^{\Gk+1})\cos{\pi (\Gk+1)\over 2h}\over 2 \sin{\pi (\Gk+1)\over 2h}}$$
which vanishes for $\Gk$ odd,  corresponding to quaternionic (half-integer spin) \rep s. 
 This result, obtained here in an explicit manner, will be recovered and generalized below for all integrable weights corresponding to irreducible representations $\kappa$ of quaternionic type.

\subsection{A case by case study}

In all cases we shall compare the results of  appendix \ref{sec:representationtypes} describing representation types for irreducible representations with the results gathered in appendix \ref{sec:automorphisms},
 that allow us to calculate the values of  the  $\Z_N$ grading 
 $\tau = \tau(\mu)$ for quaternionic representations. We shall see that for all simple Lie groups, and for quaternionic representations the quantity  $\tau$ (or at least one of the possible $\tau$'s associated with an appropriate automorphism) does not vanish. Like in section \ref{sec:proofoftheorem2} we then consider the $S_{\lambda \kappa}$ matrix elements and notice that the exponential factor appearing in  (\ref{expophase}) or in (\ref{expophasesum})  is not equal to $1$ for such representations. This shows immediately  that $\sum_\lambda S_{\lambda \kappa} = 0$ if $\kappa$ is of quaternionic type.

\subsubsection{The case $A_n \sim \su(n+1)$}

Quaternionic representations may only exist when $n+1 = 2 \, \mod \, 4$. Their highest weight $\mu$ should have Dynkin labels  that are symmetric with respect to the middle point (the position labeled $(n+1)/2$), and the middle 
Dynkin label  should be odd. 
Calculating the $N$-ality  $\tau$ of these representations (here $N=n+1$) we see immediately that only the middle term $(n+1)/2 \; \mu_{(n+1)/2}$ survives: being a product of two odd factors,  it is also odd and does not vanish modulo the even integer $n+1$.

\subsubsection{The case $B_n \sim {\rm so}(2n+1)$}

Irreps of $B_n$ are quaternionic if and only if, simultaneously,  $n=1$ or $2$ modulo $4$ and $\mu_n$ is odd.
Notice that among {\sl fundamental} irreps, only the last one (the spinorial) {\sl can be} quaternionic.
This result may be put in relation with Clifford algebra considerations since, in terms of spin groups ${\rm Spin}(d)$ with $d$ odd, quaternionic representations appear when $d$ is equal to $3$ or $5$ modulo $8$.
The $\Z_2$ grading $\tau$ (a ``2-ality'' in this case)  of a quaternionic irrep never vanishes since $\mu_n$ is odd for such representations.
	
\subsubsection{The case $C_n \sim {\rm sp}(2n) $}
Convention: the last root (to the right) is long.
Irreps are of quaternionic type whenever  $\mu_1+\mu_3+ \mu_5+\ldots+\mu_m$ is odd (where $m=n$ if $n$ is odd and $m=n-1$ if $n$ is even). But then, their $\Z_2$ grading $\tau$ is equal to $1$, and the discussion goes as before with the same conclusion.

\subsubsection{The case $D_n \sim {\rm so}(2n)$}
Convention: the  end points of the ``fork" of the Dynkin diagram 
are to the right, in positions $n-1$ and $n$.
We assume  $n \geq 3$. Remember that $D_3 \sim A_3$.
The irreps are quaternionic if and only if, simultaneously,  $n = 2  \mod \,  4$ and $\mu_{n-1}$ + $\mu_n$ is odd.
This implies that either $\mu_{n-1}$ is odd or $\mu_{n}$ is odd, but not both. 

It is not too difficult to prove that, in such a case, one of the two gradings $\tau^\prime$  or $\tau^{\prime \prime}$ associated with the two generators $\ssigma^\prime$ and $\ssigma^{\prime \prime}$ of $\Z_2 \times \Z_2$ will not vanish, but it is much simpler, and actually immediate, to use the product of these two generators (see the table in appendix  \ref{sec:automorphisms}),  with associated grading $\tau^{\prime \prime\prime}$ since it gives directly  $\tau^{\prime\prime\prime}(\mu) =2 (\mu_{n-1} + \mu_n)  \, \mod 4$, so that  $\tau^{\prime\prime\prime}(\mu)= 2 \neq 0$ for quaternionic representations.

\subsubsection{The case $E_7 $}
An irrep $\mu$ is of  quaternionic type iff $\mu_1+ \mu_3+  \mu_7$ is odd (read our convention for vertices of $E_7$ at the end of appendix \ref{sec:automorphisms}).
The center is now  $\Z_2$ and the associated grading is  $\tau(\mu)=\mu_1+ \mu_3+  \mu_7 \, \mod \, 2$.
We reach immediately the conclusion that $\tau$ does not vanish for irreps of quaternionic type.

\subsubsection{The cases $D_{odd}, G_2, F_4, E_6, E_8$}

\noindent
All the irreps of $G_2, F_4, E_8$ are self-conjugate of real type.
Not all the irreps of $E_6$ are  self-conjugate, but all self-conjugate irreps are of real type. 
The irreps of $D_{odd}$ are real or complex according as the last two components of their highest weight, but they are never quaternionic.

\noindent
Therefore, in the above cases, there is nothing else to discuss, as far as quaternionic irreps are concerned. 

\medskip
This case by case study  completes the proof of Theorem 4.


\section{The case of real representations} 
\label{sec:realrepresentations}

It may happen that $ \GS(\Gk)=\sum_\lambda S_{\lambda \Gk}$ still vanishes for some representation $\mu$ of real type. This can be the consequence of the existence of some non-trivial automorphism of the Weyl alcove
associated with a non-zero grading $\tau$, 
 but it can just be an accidental property of the chosen representation. Notice that there are no non-trivial automorphisms for $F_4$, $G_2$ and $E_8$  anyway.

\subsection{About the vanishing of $\Sigma(\kappa)$, for $\kappa$ real,  implied by automorphisms with non-zero associated grading}

Using 
 together the tables of appendices \ref{sec:automorphisms} and \ref{sec:representationtypes}, it is easy to see that, for real representations, $\tau$ is always $0$ for $A_n, C_n, E_6$ and $E_7$.
Hence, in these cases, there is no constraint on representations $\mu$ of real type coming from the existence of automorphisms, and we therefore expect that $ \GS(\Gk)$  
will be generically non-vanishing.

\smallskip
\noindent
For irreps of real type of $B_n$ and $D_n$ we find non-trivial constraints.

$B_n$.
If $n=0,3 \mod 4$, then choosing  the last component $\Gk_n$ of $\kappa$  to be  odd, leads to a non-trivial $\tau$,  so that the sum 
$\GS(\Gk)$  
vanishes.
If $n=1,2 \mod 4$ we do not find any constraint on this sum for real representations (they are such that $\Gk_n$ is even), but remember that this sum vanishes when $\Gk_n$ is odd since the representation is then quaternionic.

$D_n$ (here $n$ can be even or odd). 
Take $ \Gk$  
an irrep of real type (see table in appendix  \ref{sec:representationtypes}), then the sum $ \GS(\Gk)$
is zero as soon as one of the following three quantities
$2  \sum_{j=1, j \, odd}^{n-3} \Gk_j  + 2 \Gk_{n}$, 
$ \quad 2  \sum_{j=1, j \, odd}^{n-3}  \Gk_j  + 2 \Gk_{n-1}$, or
$2 \Gk_{n-1} + 2 \Gk_{n}$ does {\sl not} vanish modulo $4$.


\subsection{About accidental vanishing of $\Sigma(\kappa)$, for $\kappa$ real}

Notice first that vanishing properties of $\Sigma(\kappa)$ discussed so far are level independent, in the sense that they will hold for all values of the level $k$, provided $\kappa$ itself exists at the chosen level (i.e.  $\CK(\Gk) \le k$).  This is not so for the accidental vanishing cases that we discuss now.
For definiteness let us call ``accidental vanishing at level $k$'' a case where $\Sigma(\kappa)=0$ although this is not implied by any of the already known criteria, in particular $\kappa$ should be of real type and the vanishing property should not be the consequence of the existence of already discussed non-trivial automorphisms.
The very nature of the problem implies that the best we can do in this section is to mention our numerical observations. Such experiments rest on the calculation of the modular $S$ matrix, for various choices of the Lie algebra $\go$, and for relatively small values of the level.

The only accidental vanishing properties that we observed occur in the cases $F_4$ (we made tests up to level $4$) and $G_2$ (we made tests up to level $12$).
We know that all representations of these algebras are of real type, and that their Dynkin diagrams do not have automorphisms.
In both cases, we noticed nevertheless several cancellations
 of  $\Sigma(\kappa)$ (only for even levels in the case of $G_2$).
For $G_2$, we found $2$ cases at level $4$, $2$ cases  at level $6$, $5$ cases at level $8$, $6$ cases  at level $10$, $11$ cases  at level $12$.
For $F_4$, we found $2$ cases at level $3$ and $1$ case at level $4$.
These cancellations are level specific but some of them have a tendency, in some sense, to stabilize:  
indeed some representations $\kappa$ make $\Sigma$ vanish at some level but not at higher levels, 
whereas other $\kappa$, that appear at some level and make $\Sigma$ vanish,  seem to stay at higher level (shifted by $+ 2$ in the case of $G_2$). 
Admittedly we have no explanation at the moment for these observations.

This level dependence of accidental vanishing cases should be contrasted with,  for example,  a ``simple'' case  like $E_6$ (that we tested up to level $4$) where no accidental vanishing appears.
Here, at level $3$,  one finds $16$ weights that make $\Sigma$ vanish (among the $20$ integrable ones), but those $16$ are still present among the $34$ that make  $\Sigma$ vanish at level $4$ (there are $42$ integrable representations at that level).  As it was shown in previous sections these  cancellations are associated with the existence of complex irreps.


\subsection{Remark}
The type (complex, real or quaternionic) of irreps in the affine/quantum case $\goh_k$ at level $k$ is the same as the type obtained classically (ie $k \rightarrow \infty$), for irreps of the associated Lie algebra $\go$. The corresponding conditions on Dynkin labels can be found in articles or books on representation theory of Lie groups \cite{Fell}, \cite{Simon}. One can however take advantage of the finiteness of the number of simple objects in the category defined by $\go$ at level $k$ to obtain a closed formula generalizing, to this context,  the Frobenius-Schur indicator used in the theory of finite groups. Such a formula, that we recall in Appendix \ref{sec:FS} was proposed in \cite{Bantay}, see also \cite{SchauNg}, although we find more handy to use another expression (also given in Appendix \ref{sec:FS}). One can, for any chosen example, use this indicator to determine the representation type directly in terms of the $S$ and $T$ matrices, without relying on the classification of representation types for Lie algebras given in appendix  \ref{sec:representationtypes}.

\section{The case of finite groups}
 \label{sec:finitesubgroupsection} 
\def\varpi{f}

Is there an analogue of Theorem 1 true  for finite groups? 
Let $G$ be a finite group. We label its irreps 
$V_i$  by an  index $i=1,2,\cdots, r$
and its conjugacy classes $C_a$  by $a=1,2,\cdots, r$ ; $\bar \imath$ refers to the complex conjugate 
irrep of $i$.  
Let $N_{ij}^{\ \ k}$ stand for the multiplicity of irrep $k$ in $i\otimes j$. 
Do we have like in Theorem 1 
\be \sum_k N_{ij}^{\ \ k}\buildrel ? \over {=}\sum_k N_{\bar \imath j}^{\ \ k}\,. \label{theorem1?}\ee 
We first observe that (\ref{theorem1?}) is trivially true for the group $\Z_n$ 
for which the $j$-th representation is $z\mapsto z^j$, $z$ a $n$-th  root of 1, 
 $N_{ij}^{\ \ k}=\Gd_{i+j,k\mod n}$ and hence $\sum_k N_{ij}^{\ \ k}=1=\sum_k N_{\bar\imath j}^{\ \ k}$.

\begin{figure}\begin{center} 
\resizebox{11cm}{!}{\includegraphics[width=0.9\textwidth]{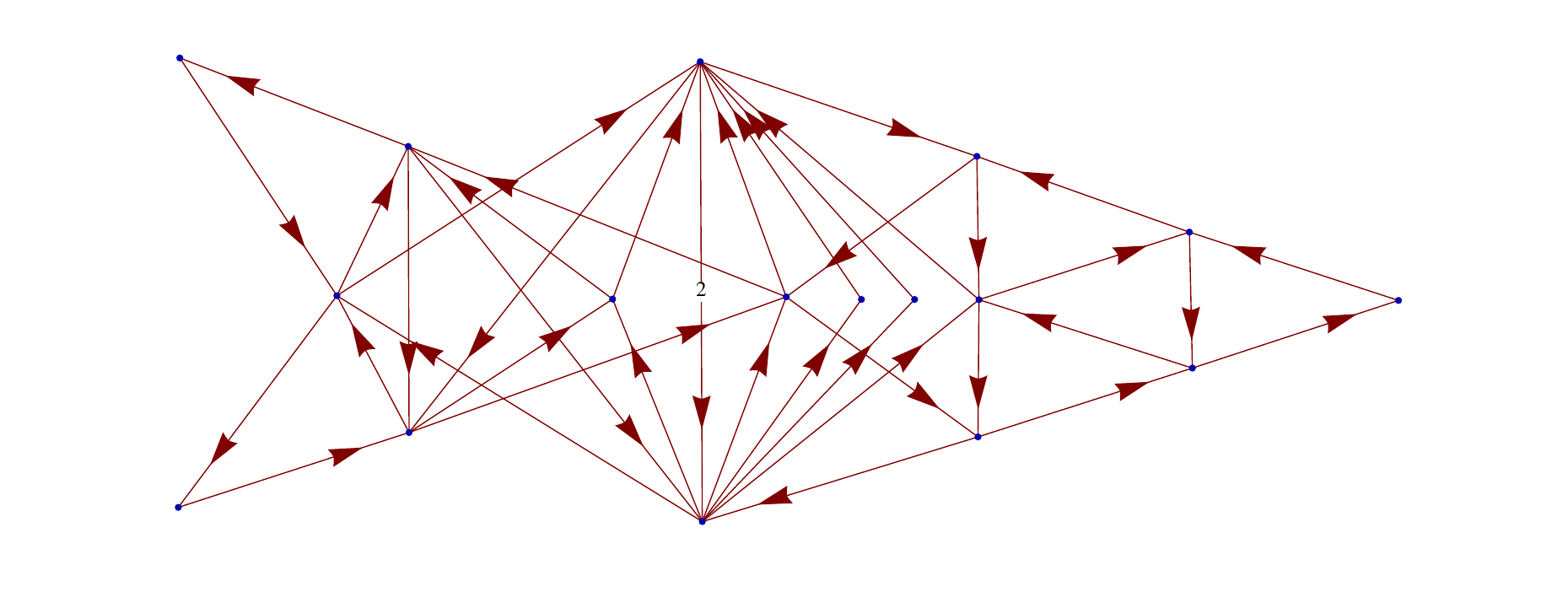}} 
\end{center}
\caption{The tensor product graph $N_\varpi$ for the subgroup $\Sigma(1080)$}
\label{subgrSU3}
\end{figure}

 To probe (\ref{theorem1?}), we have to consider less trivial groups possessing complex representations
and it is natural to look at subgroups of $\SU(3)$. 
Consider for example the subgroup of $\SU(3)$ of order 1080, 
called  $L$ 
or $\Sigma(3\times 360)$ 
in the nomenclatures\footnote{ Warning: groups $\Sigma(n)$ associated with groups $\Sigma(3 \times n)$  are subgroups of $\SU(3)/\Z_3$, not of $\SU(3)$.}
 of Yau-Yu \cite{YY} and of Fairbairn et al \cite{FFK}. It has 17 conjugacy classes and 17 irreps, 
including one of dimension 3, that we denote $\varpi$, which is the restriction of the defining representation of $\SU(3)$. 
On Fig. \ref{subgrSU3}, we display the tensor product graph $N_\varpi$, computed using the character table
given in \cite{Fly} (see also \cite{Ludl}): its vertices $i$ label the 17 irreps $V_i$, and there
are $N_{\varpi j}^{\ \ k}$ edges from $j$ to  $k$. 
A $2$ has been appended to the only (vertical) edge for which $N_{fj}^{\ \ k}=2$, all the others being equal
to 1. The graph has been drawn in such a way that complex conjugate representations are images in 
a reflection through the horizontal axis. Then Theorem 1, if true in that case, would imply that 
 the total number $\sum_k N_{fj}^{\ \ k}$ of outgoing edges from any vertex $j$ equals that  from  vertex $\bar \jmath$;
  or alternatively, that for an arbitrary vertex $k$, the number $\sum_j N_{fj}^{\ \ k}$ of incoming oriented edges 
 is equal to the number  $\sum_j N_{fk}^{\ \ j}$ of outgoing oriented edges. 
 
 It is clear on the Figure that this is not true in general, see for example the two vertices in the upper and lower middle positions.

On the other hand, we found that (\ref{theorem1?}) holds true for most subgroups of $\SU(3)$ but fails for 
some subgroups like $F=\Sigma(3\times 72)$ or $L=\Sigma(3\times 360)$. 
We could not find the criterion of validity. 

\def\bi{\bar i}
As the multiplicity $N_{ij}^{\ \ k}$ may be written as a sum over classes of characters
\be N_{ij}^{\ \ k} =\sum_a {|C_a|\over |G|} \chi_i(a) \chi_j(a) \chi_k^*(a)  \,,\ee
whose analogy with (\ref{verlinde}) is manifest, 
it is natural to wonder if Theorem 3 admits itself an analogue, whenever (\ref{theorem1?}) holds true.
In other words, do we have
\be    \sum_k \chi_k(a) \buildrel ?\over{=} 0 {\ \rm if}\  a\ne \bar a \label{sumofchi?}\ee
where $\bar a$ labels the class of the conjugates\footnote{
 Here $G$ denotes a concrete subgroup of $\SU(3)$, and complex conjugation is well defined.}
of the  elements of $C_a$.   
Just like in Sect 1, it is clear that (\ref{sumofchi?}) implies (\ref{theorem1?}), since
$\ \chi_{\bi}(a)= \chi_i(\bar a)$. And conversely, just like in Sect 4, we can prove 
that (\ref{sumofchi?}) follows from (\ref{theorem1?}).  Thus (\ref{sumofchi?}) fails 
for some of the subgroups of $\SU(3)$, like  $F=\Sigma(3\times 72)$ or $L=\Sigma(3\times 360)$.

We conclude that the validity for finite groups of (the analogues of) Theorems 1 and 3 is not to be taken  for  granted in general.

Its validity for Lie groups and affine algebras
might be an indication that the existence of the
Weyl group is an important ingredient, but this point
should be clarified. 


\section{Applications and discussion}

\subsection{Nimreps and  boundaries}

The property of the fusion algebra encapsulated in Theorems 2 and 3 has consequences 
on representations of that algebra. Particularly interesting are the 
non-negative integer valued matrix representations\footnote{It may happen that some nimreps, dubbed 
``non-physical'',  do not describe any boundary cft, or in a categorial language, any
``module-category'' for the chosen fusion category.
Unless otherwise specified, we are only interested  in the physical ones. } (``nimreps")
of the fusion algebra, namely matrices $n_\Gl$ with non-negative entries $(n_\Gl)_a^{\ b}$ satisfying 
\be \label{fusalg} n_\Gl n_\Gm=\Nh_{\Gl\Gm}^{\ \ \Gn} n_\Gn \ .\ee
They  describe the action $\lambda \, a = \sum_b (n_\Gl)_a^{\ b}  b$ of the fusion ring  on its modules and they
are known to play a role in various physical or mathematical contexts. In particular in 
boundary conformal field theory, 
$(n_\Gl)_a^{\ b}$ gives the multiplicity of \rep\ $\Gl$ for the WZW theory associated with the 
affine algebra $\goh$,  on an annulus with boundary conditions 
labelled by $a$ and $b$ \cite{BPPZ,Ca}. 
The nimreps, also known as annular matrices (see for instance \cite{CSsu4CIS}), are used, as well, in the context of topological field theories.

In general, these commuting normal matrices may be diagonalized in a common orthonormalized basis $\psi$
in the form
\be n_{\Gl a}^{\ \ b}=\sum_{\Gk \in \CE}  \psi^{(\Gk)}_a\psi^{(\Gk)*}_b{S_{\Gl\Gk}\over S_{0\Gk}} 
\label{nimrep}
\ee
with eigenvalues ${S_{\Gl\Gk}\over S_{0\Gk}} $ of the same form as those of $\N_\Gl$, but 
labelled by a subset  $\CE$  of h.w.   $\Gk$ called exponents.
The $\psi$'s enjoy  conjugacy properties similar to those of the $S$ matrix, 
in particular  
\be \label{conjpsi} \psi^{(\Gk)*}_b =\psi^{(\bar\Gk)}_b 
\,. \ee
The subset of exponents  is closed under conjugacy, so that the above equation implies immediately 
$(n_{\bar \Gl})_a^{\ b} = (n_\Gl)_b^{\ a}$ \ie $n_{\bar \lambda} = n_\lambda^T$.
 The matrices $n_\Gl$ may be regarded as adjacency matrices of a collection of graphs, with vertices labelled
by indices $a,b,\cdots$ refering to a particular  basis $Vert$ of the chosen module. 

Automorphisms $\zeta$ of the underlying affine Lie algebra at level $k$ act both on the fusion ring and on its associated modules. They are often called {\sl symmetries}.
For instance, the transformation $\lambda \mapsto k -\lambda$ is a symmetry of the fusion ring of  $\SU(2)$ at level $k$.
It is enough to know the action of the generator(s) described in Appendix B.
On the fusion ring, we have $\zeta(\lambda \mu) = \zeta(\lambda) \mu = \zeta(\mu) \lambda$, in particular
$\zeta(\lambda) = \zeta(\one) \lambda$ where $\one = (0,0,\ldots, 0)$ labels the trivial representation of the Lie algebra.
In terms of fusion matrices the symmetry property reads $\hat N_{\Gl \zeta(\Gm)}^{\ \ \zeta(\Gn)} = \hat N_{\Gl \Gm}^{\ \ \Gn}$.
On a module, the action is specified by setting $\zeta(a) = \zeta(\one) \, a$ for all $a \in Vert$. One obtains immediately 
$\zeta (\lambda a) = \zeta(\one) \, \lambda \, a = \lambda \,  \zeta(\one) a = \lambda \zeta(a)$. 
We denote by the same symbol $P$ the matrices describing multiplication by  $\zeta(\one)$ both in the fusion ring and in the module, \ie $P=N_{\zeta(\one)}$ or $P=n_{\zeta(\one)}$.
Obviously  $\hat N_{\zeta(\Gl)} = \hat N_\Gl \, P$ and $n_{\zeta(\Gl)} = n_\Gl \, P$.
Denoting by the same symbol $X$ the two matrices\footnote{These ``path matrices'' $X$ are discussed in section 8.4}   $\sum_\lambda \hat N_{\lambda}$ and $\sum_\lambda \hat n_{\lambda}$,
one obtains immediately $X P = X$ since the action of $\zeta$ is one-to-one. \\

A complex {\sl conjugation} in the module is  an involution\footnote{When conjugation in the fusion ring itself is trivial, there is no need to introduce this concept.} $a \mapsto  \bar a$ such that  $\overline{\lambda a } = \bar{\lambda}  \bar{a}$.
 If the basis  $Vert$ used to label the nimreps is stable as a set under transformations $a \mapsto \zeta(a)$ and $a \mapsto  \bar a$, the previous conditions read respectively
 $n_{\lambda \zeta(a)}^{\ \ \ \zeta(b)} =  n_{\lambda a}^{\ \ \ b}$ and
$ n_{{\bar \lambda}{\bar a}}^{\ \ {\bar b}} = n_{\lambda \, a}^{\ \ \ b} \quad  \text {for all}  \; \lambda, a, b$.  
One can always define a matrix $C$ with $C^2=1$ such that  $n_{\bar \lambda}  =C \,  n_\lambda \,  C$.
From a given conjugation in a module one can obtain another one by composing it with a symmetry. 
Usually an involution $a \mapsto \bar a$ is determined, up to symmetry, from the known conjugacy properties of the set of exponents, but there may nevertheless remain
an ambiguity when some exponents have multiplicity higher than $1$. The ambiguity in the definition of $C$ reflects a potential ambiguity in the definition of the diagonalizing $\psi$ matrix because $C$ can be defined as $\psi^T \, \psi$. Notice that the matrix $\psi \, \psi^T$ gives the restriction of the known conjugation matrix of the Lie algebra at level $k$ to the corresponding set of exponents.


This discussion applies in particular to  the nimreps of the $\widehat{su}$(2) algebra,  which are 
in one-to-one correspondence with the ADE Dynkin diagrams (plus the  
``tadpole" diagrams\footnote{that actually describe non-physical nimreps.} $T_n=A_{2n}/\Z_2$). All irreps at level $k$ are self-conjugate but there is a non-trivial involution $P$ on the $A_n$ diagrams, that induces a non-trivial involution on the $D_{n=2s+1}$ and $E_6$ diagrams. Here $a\mapsto \zeta(a)$ is just the $\Z_2$ symmetry  of the Dynkin diagram. 
For the $D_{even}$ diagrams, the matrix $P$ is trivial, although we still have a non-trivial geometrical symmetry that exchanges the two branches of the fork, {\it i.e.} a graph automorphism\footnote{Graph automorphisms are permutations $\pi$ on vertices of a graph such
that for all pairs of vertices, $(\pi(a),\pi(b))$ is an edge iff $(a,b)$ is an edge.}.
Notice that symmetries of a module structure over the fusion ring, as defined in the text, give rise to automorphisms of fusion graphs, but there may be more of the latter.
In the case of nimreps of the $\widehat{su}$(3) algebra, the various diagrams exhibit several interesting geometrical symmetries, but besides the diagrams of type ${\mathcal A}$ themselves, 
only the exceptional diagram with self-fusion at level $5$ and  the diagrams of the conjugated Dstar family,  when the level is not $0$ modulo $3$, admit a non-trivial matrix $P$ inherited from the $Z_3$ symmetry of the corresponding fusion algebra.  For all these cases  $X \, P = X$ and $P$ is non-trivial.

In the case of the $\SU(2)$ WZW model,   the equation $X = X P$ means 
that the total number of \rep s, i.e. of primary fields contributing
to the annulus partition function in the presence of boundary conditions $a$ and $b$ 
is the same as with b.c. $a$ and $\zeta(b)$. 
This extends to the minimal 
$c<1$ cft's, that are constructed as cosets of the $\su(2)$ theories. They are
classified by a pair $(A_{h'-1},G)$ of Dynkin diagrams, $G$ of ADE type and of Coxeter number $h$. 
Their boundary conditions are classified by pairs $(\rho,a)$ with $\rho=1,\cdots, h'-1$ and $a$ a vertex
of $G$ \cite{BPPZ}. 
Take one of the cases $G=A,D_{{\rm odd}}, E_6$. 
The multiplicity $n_{rs;\, (\Gr,a)}^{\qquad (\Gr',b)}$ of the  $(r,s)$ primary field in the 
annulus partition function with boundary conditions $(\Gr,a)$ on one side and $(\Gr',b)$ on the other is
not invariant under the symmetry $b\mapsto \zeta(b)$, 
but the total multiplicity $\sum_{s} n_{rs;\, (\Gr,a)}^{\qquad (\Gr',b)}$ is, 
as one may check for example on the explicit formulae of \cite{SalBau} in the case of $E_6$.

In general, conjugacy properties of the $\psi$'s imply  (or are implied by) conjugacy properties of the $n$'s,
but Theorem 4 implies stronger properties for  {\it sums} of the $n$'s.  
Following  steps similar to those in (\ref{imply})  in Sect 1 and 
making use of $\psi^{(\Gk)}_b=\psi^{(\Gk)}_{\bar b}$ for real $\kappa$, one finds $X = X \, C$. Indeed,
\be \label{sum-nimrep}  
\sum_\Gl n_\Gl= (\sum_\Gl n_\Gl)^T= C \sum_\Gl n_\Gl =\sum_\Gl n_\Gl C\ 
\Longleftrightarrow\ 
\sum_\Gl  n_{\Gl a}^{\ \ b}=\sum_\Gl  n_{\Gl a}^{\ \ \bar b}
\ee

 Like for $S$ itself, we have observed, in many cases,  intriguing sum rules concerning the matrix $\psi=(\psi_a^{(\kappa)})$, involving summations either  over the exponents, or 
 over the label $a$. We hope to return to this analysis in a later work.

\ommit{
\blue On peut etre finalement assez precis a cet endroit... Mais la description prendra au moins une demi-page... Alors ? \colend
\blue Je n'ai pas insere la discussion concernant les phases (version precedente). Les resultats etaient corrects, bien sur, mais ce sont des commentaires qui prennent de la place et je ne vois pas bien pourquoi on devrait necessairement les mentionner (?). \colend
\blue Ce qu'il y a de ``special'' avec le E5 de SU3 ? Il semble que ce soit le seul cas de la famille SU3 (mis \`a part les diagrammes de type A) o\`u la matrice $P$ n'est pas triviale.
Son cube est ici egal \`a 1. C'est exactement l'analogue de la discussion faite pour le Dodd de SU2.  Je dois verifier mais je crois bien.. \colend
}
\ommit{
The property of the fusion algebra encapsulated in Theorems 2 and 3 has consequences 
on representations of that algebra. Particularly interesting are the 
non-negative integer valued matrix representations\footnote{It may happen that some nimreps, dubbed 
``non-physical'',  do not describe any boundary cft, or in a categorial language, any
``module-category'' for the chosen fusion category.
Unless otherwise specified, we are only interested  in the physical ones. } (``nimreps")
of the fusion algebra, namely matrices $n_\Gl$ with non-negative entries $(n_\Gl)_a^{\ b}$ satisfying 
\be \label{fusalg} n_\Gl n_\Gm=\Nh_{\Gl\Gm}^{\ \ \Gn} n_\Gn \ .\ee
They  describe the action of the fusion ring  on its modules and they
are known to play a role in various physical or mathematical contexts. In particular in 
boundary conformal field theory, 
$(n_\Gl)_a^{\ b}$ gives the multiplicity of \rep\ $\Gl$ for the WZW theory associated with the 
affine algebra $\goh$,  on an annulus with boundary conditions 
labelled by $a$ and $b$ \cite{BPPZ,Ca}. 
The nimreps, also known as annular matrices (see for instance \cite{CSsu4CIS}), are used, as well, in the context of topological field theories.
 }
\ommit{
In general, these commuting normal matrices may be diagonalized in a common orthonormalized basis $\psi$
in the form
\be n_{\Gl a}^{\ \ b}=\sum_{\Gk \in \CE}  \psi^{(\Gk)}_a\psi^{(\Gk)*}_b{S_{\Gl\Gk}\over S_{0\Gk}} 
\label{nimrep}
\ee
with eigenvalues ${S_{\Gl\Gk}\over S_{0\Gk}} $ of the same form as those of $\N_\Gl$, but 
labelled by a subset  $\CE$  of h.w.   $\Gk$ called exponents.
 The matrices $n_\Gl$ may be 
regarded as adjacency matrices of a collection of graphs, with vertices labelled
by the indices $a,b,\cdots$. 
The $\psi$'s enjoy  conjugacy properties similar to those of the $S$ matrix, 
in particular  
\be \label{conjpsi} \psi^{(\Gk)*}_b =\psi^{(\bar\Gk)}_b 
\,. \ee
One   has often an  
involution $b\to \bar b$ such that if $C$ is the matrix $C_{ab}=\Gd_{a\bar b}$,
$C^2=I$, $C n_\Gl C=(n_{\bar\Gl})^T$, and $\psi^{(\Gk)}_{\bar a}=
\eta_\Gk \psi^{(\bar\Gk)}_a$, i.e.
\be C\psi^{(\Gk)}=\eta_\Gk \psi^{(\bar\Gk)} \ ,\ee
with $|\eta_\Gk|=1$. Strictly speaking this has
to be elaborated in cases where some exponents have multiplicity higher than 1, but this point
is beyond the scope of this paper and we hope to return to it later.
For real or quaternionic weights $\kappa=\bar \kappa$, the $\psi$'s 
are real according to (\ref{conjpsi}) and the phase $\eta_k=\pm 1$.  In all 
nimreps of $\widehat{su}$(2), 
 we have observed that 
the real \rep s have $\eta_\kappa=1$ and the quaternionic ones have $\eta_\kappa=-1$. 
}
\ommit{
In general these conjugacy properties of the $\psi$'s imply  (or are implied by) symmetries of the $n$'s,
 such as  $C n_\Gl C=n_\Gl^T =n_{\bar \Gl}$
 or $  n_{\Gl \bar a}^{\ \ \, \bar b}= n_{\Gl b}^{\ \  a}  = n_{\bar\Gl a}^{\ \  b}$.
Theorem 4 implies further symmetries of {\it sums} of the $n$'s.  
Following  steps similar to those in (\ref{imply})  in Sect 1 and 
making use of $\psi^{(\Gk)}_b=\psi^{(\Gk)}_{\bar b}$ for real $\kappa$, one finds
\be\label{sumn}
\sum_\Gl n_\Gl= (\sum_\Gl n_\Gl)^T= C \sum_\Gl n_\Gl =\sum_\Gl n_\Gl C\ ,
\ee
or $\sum_\Gl n_{\Gl a}^{\ \  b}=\sum_\Gl n_{\Gl b}^{\ \  a}=\sum_\Gl n_{\Gl \bar a}^{\ \  b} =\sum_\Gl n_{\Gl  a}^{\ \ \, \bar b}$.
}
\ommit{
This applies in particular to  the nimreps of the $\widehat{su}$(2) algebra,  which are 
in one-to-one correspondence with the ADE Dynkin diagrams (plus the  
``tadpole" diagrams\footnote{that actually describe non-physical nimreps.} $T_n=A_{2n}/\Z_2$).
There exists  a  nontrivial involution $C$ for 
the $A_n$, $D_{n=2s+1}$ and $E_6$ diagrams, where  $a\mapsto \bar a$ is
just the $\Z_2$ symmetry  of the Dynkin diagram.
Moreover as mentionned above the phase $\eta_\Gk$ is then found
to be equal to $\pm 1$, depending on whether the \rep\ $\Gk$ of $\su(2)$ is
real or quaternionic  (integer or half-integer spin). Thus according to (\ref{sumn})
\be \sum_\Gl  n_{\Gl a}^{\ \ b}=\sum_\Gl  n_{\Gl a}^{\ \ \bar b}\ , \label{sum-nimrep} \ee
}
\ommit{
In words, (\ref{sum-nimrep}) means that 
the total number of \rep s, i.e. of primary fields of the $\SU(2)$ WZW model, contributing
to the annulus partition function in the presence of boundary conditions $a$ and $b$ 
is the same as with b.c. $a$ and $\bar b$. 
This extends to the minimal 
$c<1$ cft's, that are constructed as cosets of the $\su(2)$ theories. They are
classified by a pair $(A_{h'-1},G)$ of Dynkin diagrams, $G$ of ADE type and of Coxeter number $h$. 
Their boundary conditions are classified by pairs $(\rho,a)$ with $\rho=1,\cdots, h'-1$ and $a$ a vertex
of $G$ \cite{BPPZ}. 
Take one of the cases $G=A,D_{{\rm odd}}, E_6$. 
The multiplicity $n_{rs;\, (\Gr,a)}^{\qquad (\Gr',b)}$ of the  $(r,s)$ primary field in the 
annulus partition function with boundary conditions $(\Gr,a)$ on one side and $(\Gr',b)$ on the other is
not invariant under the conjugation $b\mapsto \bar b$, 
but the total multiplicity $\sum_{s} n_{rs;\, (\Gr,a)}^{\qquad (\Gr',b)}$ is, 
as one may check for example on the explicit formulae of \cite{SalBau} in the case of $E_6$.
}
\ommit{
 Like for $S$ itself, we have observed, in many cases,  intriguing sum rules concerning the matrix $\psi=(\psi_a^{(\kappa)})$, involving summations either  over the exponents, or 
 over the label $a$. We hope to return to this analysis in a later work.
}

\subsection{ Integrable S-matrices}

The nimreps of the previous section have appeared in a different context, that of 
S-matrices of integrable 2-d field theories.
In the study of affine Toda theories or of other integrable 2-d theories 
based on a simply laced algebra, Braden, Corrigan, Dorey and   Sasaki\cite{BCDS}
were led to expressions, proved or conjectured, of their scattering $\CS$-matrix. Typically
the particles of those theories are in one-to-one correspondence with the vertices of
ADE-Dynkin diagrams. 

The  $\CS_{ab}$ matrix describing the scattering of particles $a$ and $b$ is a function of the relative 
rapidity $\theta=\theta_a-\theta_b$ and satisfies the contraints of\\
-- unitarity $\CS_{ab}(\theta)\CS_{ab}(-\theta)=I$ and \\ 
-- crossing    $\CS_{ab}(\theta)
=
\CS_{ b\,  \zeta(a)}(i\pi - \theta)$\\
which imply that  $S_{ab}$ is $2\pi i $ periodic. Its analytic structure may be investigated 
in the strip $0\le \Im m\, \theta < \pi$, from which the whole period may be recovered. 
One finds that in that strip, it has poles at 
$\theta= \vartheta_\ell:=\ell {i\pi\over h}$, with $h$ the Coxeter number of the ADE diagram 
and $\ell=1,\cdots, h-1$. Quite amazingly\cite{Dorey1,Dorey2}, 
the multiplicity of the pole at $\vartheta_\ell$ turns out to be  $n_{\ell-2\, a}^{\qquad \! b} 
+n_{\ell \, a}^{\ \ \, b}$, where by convention $n_{-1}=n_{h-1}=0$.

In that context, the identity (\ref{sum-nimrep}), rewritten here as 
$\sum_\Gl n_{\Gl a}^{\ \ b} =\sum_\Gl 
n_{\Gl\zeta(b)}^{\ \  \  \ a}$ 
because of the symmetry of the $n$ matrices in this case, 
expresses that the total number of poles of $\CS_{ab}$ and of 
$\CS_{\zeta(b) a}$ are
equal, in accordance with the crossing relation above\footnote{We are very grateful to Patrick
Dorey for refreshing our memory and for this nice observation.
}. 


\subsection{Sum rules for character polynomials}
\label{charpol}
Call $\chi(\lambda)=\chi(\lambda; t_1,t_2,\ldots t_n)$ the classical character polynomial of the Lie group $G$, associated with an irreducible representation defined by its highest weight $\lambda$. It encodes the weight system of $\lambda$ :  each weight $\ell \in [\lambda]$ occurring with multiplicity $a$ in this weight system gives a Laurent
 monomial $a\, t_1^{\ell_1} t_2^{\ell_2} \ldots t_n^{\ell_n}$ in $\chi(\lambda)$. Here $(\ell_1, \ell_2, \ldots, \ell_n)$ are the Dynkin labels  (that can be positive or negative or zero) of $\ell$. Evaluation {\sl at level k}  of such a monomial on a weight $\mu$ is, by definition
$a\, \exp[2 i \pi/(h^{\vee}+k) \; \langle \ell_1 \omega_1 + \ell_2 \omega_2 + \ldots + \ell_n \omega_n, \mu  \rangle]$, where $\omega_i$ are the fundamental weights, and it is extended to arbitrary Laurent polynomials by linearity. The obtained value is denoted $\chi(\lambda)[\mu]$. 
Assuming that $\lambda$ and $\mu$ are two irreducible representations of $G$ existing  at level $k$, one obtains, from the Kac-Peterson formula, the following relation between the matrix elements of  $S$  and the (classical) character polynomial: 
\be \label{Schi}
S_{\lambda \mu} / S_{00} =  {\dim}_q(\mu) \; \chi(\lambda)[\mu + \rho]\,.
\ee
The quantum dimension of $\mu$ is obtained as $$\dim_q(\mu)= S_{\mu 0} / S_{00} =  \chi(\mu)[ \rho] = \chi(\mu; q^{2 \rho^1},q^{2 \rho^2} ,\ldots q^{2 \rho^r})$$
where $(\rho^j)$ are the components of the Weyl vector on the base of simple coroots (Kac labels), and $q={\exp}({ i \pi}/({h^{\vee}+k}))$.
The previous relation for $S_{\lambda \mu}$ looks asymmetrical, but since $S$ is symmetric, it implies  
$$\dim_q(\mu) \, \chi(\lambda)[\mu+ \rho] = \dim_q(\lambda) \, \chi(\mu)[\lambda+ \rho]$$
\noindent
Now, every sum rule for $S$ (Theorem 3 or Theorem 4) leads immediately to a corresponding identity for the classical character polynomial. Using the symmetry property of $S$, the quantum dimension $\dim_q(\lambda)$ can be factored out, and we obtain the following property:

\begin{center}
Call $X$ {the Laurent polynomial} $X(t_1,t_2,\ldots, t_n) \,  = \sum_{\mu \, \text{with} \,  \langle \theta, \mu \rangle \le k}\chi(\mu),$ \\ then
$X[\lambda+ \rho] = 0 \; \text {if} \;  \lambda \; $ is of complex or quaternionic type.
\end{center}

\noindent
 
Example: The  character polynomials for the 6 irreps of $\su(3)$ at level $2$, of highest weights
 $\{(0,0),(1,0),(0,1),(2,0),(1,1),(0,2)\}$ and of classical dimensions $\{1,3,3,6,8,6\}$, are:
 {\small
\begin{eqnarray*}
{}&{}&
1,\; 
\frac{ {t_2}}{ {t_1}}+ {t_1}+\frac{1}{ {t_2}},
\; \frac{ {t_1}}{ {t_2}}+\frac{1}{ {t_1}}+ {t_2},\; 
\frac{ {t_2}^2}{ {t_1}^2}+ {t_1}^2+\frac{ {t_1}}{ {t_2}}+\frac{1}{ {t_1}}+\frac{1}{ {t_2}^2}+ {t_2},\; \\
{}&{}&
 \frac{ {t_1}^2}{ {t_2}}+\frac{ {t_2}}{ {t_1}^2}+\frac{ {t_1}}{ {t_2}^2}+\frac{ {t_2}^2}{ {t_1}}+ {t_1}  {t_2}+\frac{1}{ {t_1}  {t_2}}+2, \; 
 \frac{ {t_1}^2}{ {t_2}^2}+\frac{1}{ {t_1}^2}+\frac{ {t_2}}{ {t_1}}+ {t_1}+ {t_2}^2+\frac{1}{ {t_2}} \\
\end{eqnarray*}
}
The polynomial $X(t_1,t_2)$ is 
{\small
\be\label{su32}
3+\frac{1}{ {t_1}^2}+\frac{2}{ {t_1}}+2  {t_1}+ {t_1}^2+\frac{1}{ {t_2}^2}+\frac{ {t_1}}{ {t_2}^2}+\frac{ {t_1}^2}{ {t_2}^2}+\frac{2}{ {t_2}}+\frac{1}{ {t_1}  {t_2}}+\frac{2  {t_1}}{ {t_2}}+\frac{ {t_1}^2}{ {t_2}}+2  {t_2}+\frac{ {t_2}}{ {t_1}^2}+\frac{2  {t_2}}{ {t_1}}+ {t_1}  {t_2}+ {t_2}^2+\frac{ {t_2}^2}{ {t_1}^2}+\frac{ {t_2}^2}{ {t_1}}
 \,. 
\ee}
Its evaluation on the six h.w., using $q=\exp (i \pi/5)$,  gives  $\left\{\frac{3}{2} \left(3+\sqrt{5}\right), 0,0,0,
\frac{3}{2} \left(3-\sqrt{5}\right),0\right\}$.

 
\subsection{On the path matrix $X$ and its spectral properties}
 \label{pathX}
  We use fusion matrices
defined as $\hat N_\lambda = (\hat N_{\lambda \mu}^{\ \ \nu})$.
Using standard equalities $\hat N_{\lambda \mu}^{\ \ \nu}  = \hat N_{\mu \lambda}^{\ \ \nu}$,  $\hat N_{\lambda \mu}^{\ \ \nu}  = \hat N_{\overline \lambda \overline \mu}^{\ \ \overline \nu}$, $\hat N_{\lambda \mu}^{\ \ \nu}  = \hat N_{\overline \lambda \nu}^{\ \ \mu}$, and the conjugation matrix $C$  introduced in sect. 1,
 with components $C_{\mu \nu} = \delta_{\mu \overline \nu}$, we have 
 $\hat N_{\overline \lambda} = C \hat N_{\lambda} C$.  
 Define the matrix $X = \sum_\lambda \hat N_\lambda$, dubbed ``path matrix" for reasons explained below.
 From the corresponding property for $\hat N_\lambda$  one obtains immediately $X = C X C$.
 The sum rule described by theorem 1 tells us that we can actually drop one of the two conjugation matrices in this equation. In other words, the equation $X = C X = C X$ holds.
 More generally, for any chosen module (nimrep) over the fusion algebra, one can define a path matrix $X = \sum_\lambda n_\lambda$ that enjoys similar properties.

 There exist several interpretations of fusion coefficients (more generally of coefficients of nimreps) in terms of combinatorial  constructions associated with fusion graphs: essential paths \cite{Ocneanu:Bariloche} (or generalizations of the latter), admissible triangles \cite{Kauffman:book}, (generalized) preprojective algebras or quivers\cite{MOV}, and they can also be used to define interesting weak Hopf algebras \cite{Ocneanu:paths, PetkovaZuber:Oc, TrincheroDTACoquereaux6J}. 
The translation of the sum rules involving the fusion coefficients (or those of the nimreps) into these different languages and points of view is left as an exercise to the reader. 
  In the first  combinatorial interpretation, the sum $\sum_{\nu \rho} \hat N_{\mu \nu}^{\ \ \rho}$ (or $\sum_{a b} n_{\mu a}^{\ \ b}$ for the nimreps) gives the dimension of the space of essential paths with fixed length $\mu$, and matrix elements $X_{\nu \rho}$ (or $X_{a b}$ for the nimreps) gives the dimension of the space of essential paths of arbitrary length, but with fixed origin and extremity.  This explains the name ``path matrix'' given to $X$.

It is sometimes useful to consider, instead of $S$,  the fusion character table  $\chi = (\chi_{\mu \nu})$ with $\chi_{\mu \nu} = S_{\mu \nu}/S_{0 \nu}$. 
The columns of that matrix are made of eigenvectors common to all fusion matrices (the first column giving the quantum dimensions of irreps), and the line labelled $\mu$ gives the corresponding eigenvalues for the fusion matrices $N_\mu$ (the first line being $1\ldots 1$).  The matrix $\chi$, in contradistinction to $S$,  is not symmetric.
Theorems 3 and 4 imply: $\sum_{\mu} \chi_{\mu \nu} = 0$ whenever $\nu$ is complex or quaternionic.

$\blacklozenge \,1$
As all the $\hat N_\Gm$ are diagonal in the same basis provided by the column-vectors $S[\Gn]$ of $S$ with eigenvalues
$S_{\Gm\Gn}/S_{0\Gn}$, see (\ref{verlinde}),  their sum $X$ has in the same basis the eigenvalues  $\sum_\Gm \chi_{\mu \nu}$. 
The only possible non-zero eigenvalues of the path matrix  $X$ therefore correspond to irreps of real type.
Example (continuation of (\ref{su32})): The Lie algebra $\su(3)$ at level $2$ has $6$ irreps, two of them being of real type (those of highest weight $(0,0)$ and $(1,1)$),  
the $6$th degree characteristic polynomial of the corresponding path matrix $X$ has therefore only two non vanishing roots.

$\blacklozenge \,2$
From Verlinde formula it is easy to show that 
$\sum_{\mu^\prime} \chi_{\mu \mu^\prime}  \,  \chi_{\nu \mu^\prime }  = \Tr(\hat N_\mu \hat N_\nu)\,.  $ 
In particular 
$\Tr(\hat N_\mu) = \ \sum_\nu \chi_{\mu \nu}$. 
This sum over eigenvalues of a fusion matrix is automatically an integer.\\
Warning:  The numbers $\sum_\mu \chi_{\mu \nu}$ obtained previously as eigenvalues of $X$ are usually not  integers.

$\blacklozenge \, 3$
From the relation  (\ref{Schi}) between the $S$ matrix and the classical character polynomials we obtain:
 $$ \chi_{\mu \nu} = \chi(\mu)[\nu + \rho]\,.$$
Using the final result of section \ref{charpol}, the eigenvalues of the path matrix $X$, in particular its $0$ eigenvalues, can be obtained from the evaluation of the Laurent polynomial (also called  $X$ there, on purpose), 
on the irreps that exist at the chosen level.
 \\
 Example (continuation). The path matrix of $\su(3)$ at level $2$ is easily found to be 
 {\tiny
 $
 \left(
\begin{array}{cccccc}
 1 & 1 & 1 & 1 & 1 & 1 \\
 1 & 2 & 2 & 1 & 2 & 1 \\
 1 & 2 & 2 & 1 & 2 & 1 \\
 1 & 1 & 1 & 1 & 1 & 1 \\
 1 & 2 & 2 & 1 & 2 & 1 \\
 1 & 1 & 1 & 1 & 1 & 1
\end{array}
\right)
.$
}
One can check that its non-zero eigenvalues are the two non-zero values obtained at the end of section \ref{charpol}.
 
$\blacklozenge \, 4$
Call $\frak s_1 = \sum_\mu \dim_q (\mu)$ and $\frak s_2 = \sum_\mu \dim_q (\mu)^2$.  In section 1 we  defined $\Sigma(\mu) = \sum_{\lambda} S_{\lambda\mu} =  \dim_q(\mu)\,  S_{00} \, \sum_{\lambda} \chi_{\lambda \mu}$. Using the standard result $\frak s_2 = 1/S_{00}^2$, one finds $\Sigma(\mu)=  \dim_q(\mu) \, ( \sum_{\lambda} \chi_{\lambda \mu})/  {\sqrt{\frak s_2}}$.
Since  $\dim_q(\lambda) = \chi_{\lambda 0}$, we obtain in particular $\Sigma(0) =  {\frak s_1} / {\sqrt{\frak s_2}}$.

Example (continuation). For an irrep $\lambda=(\lambda_1, \lambda_2)$ of $\su(3)$  we can use the standard formula  $\dim_q(\lambda) = {\left(\lambda _1+1\right)_q \left(\lambda _2+1\right)_q \left(\lambda _1+\lambda_2+2\right)_q}/{1_q^2 2_q}$, where $n_q = {(q^n-q^{-n})}/{(q-q^{-1})}$. By summing quantum dimensions (or their squares) over the  Weyl alcove of level $2$, we recover $\frak s_1 =  \frac{3}{2} \left(3+\sqrt{5}\right)$, that we have already obtained as the evaluation of the Laurent polynomial $X(t_1,t_2)$ on the $0$ weight, or as one of the two non-zero eigenvalues of the path matrix $X$, and calculate $\frak s_2 = \frac{3}{2} \left(5+\sqrt{5}\right)$. One finds $\Sigma(0)=\sqrt{3+\frac{6}{\sqrt{5}}}$.


\subsection{Final comments}


Admittedly our proofs of Theorems 1-4 lack conciseness and  more direct and conceptual proofs would 
be highly desirable.
For example it is natural to wonder 
if  there is a direct proof of Theorems 3 and 4, based on Galois arguments or some other hidden symmetry 
of the $S$ matrix.  If so, the proofs of Theorems 2 first
(through Verlinde formula) and 1 then (through the large $k$ limit) would follow.

Another tantalizing option would be to use Steinberg formula.
Steinberg formula for tensor multiplicities reads $N_{\lambda \mu}^{\ \ \nu} = \sum_{v,w \in W} \, {\rm sign}(vw)\,  {\cal P}( v \cdot \lambda  + w \cdot \mu - \nu)$ where $v \cdot \lambda = v[\lambda+\rho] - \rho$ is the Weyl shifted action, and  ${\cal P}$ is the Kostant partition function, which gives the number of ways one can represent a weight as 
 an integral non-negative combination of positive roots.
The highest weight of the conjugate of an irrep is the negative of the lowest weight of that irrep.
The lowest weight is obtained from the highest weight by the action of the longuest element\footnote{In all cases but $A_{even}$,  $w_0 = {\frak c}^{h/2}$ where ${\frak c}$ is a bipartite Coxeter element.} $w_0$ of the Weyl group. In other words $\overline \lambda = - w_0 [\lambda]$.\\
Our sum rule for tensor multiplicities therefore leads  to various identities involving ${\cal P}$ and $w_0$. 
Conversely, a direct proof of such identities would provide a shorter derivation of theorem 1.

\appendix
\section{The case of $E_6$}

\def\short#1{#1}  

\subsection{Sums of multiplicities for tensor products $\omega_{2,4} \otimes \mu$ of $E_6$}
\label{appA1}

\def\CNnr{\psi^{\geq 0}} 
\def\CNi{\phi}
\def\CNnr{\phi^{\geq 0}}
\def\CNiiu{\phi_{+}^{0}}
\def\CNr{\psi^{\leq 0}}
\def\CNiid{\psi_{-}^{0}}
\def\CNiii{\psi}

The detailed discussion of the 
tensor product of a \rep\ of highest weight $\mu$ by one of the fundamental representations 
$\Go_2$ or $\Go_4$ of $E_6$ offers a good illustration of the three cases (i), (ii), (iii) presented in Section 
\ref{sec:proofoftheorem1}, and is anyway a mandatory step for the completion of our proof 
of Theorem~1.  The aim of this appendix is to show how 
 the cardinalities of the two classes (i) and (iii)  and the total multiplicity 
 may be proved to be the same for $\Go_2$ and $\Go_4$.

For a given $\mu$, and $\Gl'$ one of the weights of the weight system $[\Go_2]$, 
we denote as before $\Gs=\Gl'+\Gm+\Gr$. \\
Call $\CNnr$ the number of weights $\Gs$ (counted with multiplicity) that have non-negative
Dynkin labels. Those weights need not be Weyl reflected in the Racah--Speiser algorithm.
Some of them, however, may lie on a wall of the  fundamental Weyl chamber. Call $\CNiiu$ 
the  number of the latter. The class (i) of weights with only positive Dynkin labels has thus cardinality 
$\CNi=\CNnr-\CNiiu$.
\\
If one of the  labels of $\Gs$ is negative, we shall show below that a single Weyl reflection brings  
it back to the fundamental Weyl chamber, {\it including its walls}. Call $\CNr$ the cardinality of that 
class, and $\CNiid$  
the number of the reflected weights that lie on a wall of the fundamental Weyl chamber. The 
class (iii) of weights that contribute with a minus sign to the total multiplicity has cardinality
$\CNiii=\CNr-\CNiid$.\\
The total multiplicity is finally 
$\sum_\nu N_{\Go_{2}\Gm}^{\quad \nu} = \CNi-\CNiii=\CNnr-\CNr-
\CNiiu+\CNiid$.
Note that  $\CNnr+\CNr=\CNi+(\CNiiu+\CNiid) + \CNiii=351$, the dimension of the $\Go_2$  and 
$\Go_4$ \rep s. 
All these numbers depend on the weight $\mu$. What we want to prove is that 
for a given $\mu$, $\sum_\nu N_{\Go_{\!\!{\atop 2}}\Gm}^{\ \ \ \nu}=\sum_\nu N_{\Go_{\!\!{\atop 4}}\Gm}^{\ \ \ \nu}$.
In fact we shall establish that the numbers $\CNi,\ \CNiii$ are the same for $\Go_2$ and 
$\Go_4$.

\def\non{\,{\approx \!\!\!\!\!\slash}\,}

$\bullet$ Let us first examine the $\CNr$ weights $\Gs$ that have a negative Dynkin label. As the weights $\Gl'$ of the $[\Go_2]$
or $[\Go_4]$ systems have their labels equal to $0, \pm 1,\pm2$, and at most one label equal to $-2$, 
$ \Gs_i= \Gl'_i + \Gm_i +\Gr_i=\Gl'_i + \Gm_i +1\ge -1$;  for a given $\Gs$, at most one Dynkin label $\Gs_j$
equals $-1$, and this requires $\Gl'_j=-2$ and $\Gm_j=0$. Conversely for each $j$ such that $\Gm_j=0$, 
there are as many $\Gl'$ fulfilling the above condition as there are weights $\Gl'$ with $\Gl'_j=-2$. Both in
the $[\Go_2]$ and the $[\Go_4]$ systems, this number is 15. Thus $\CNr= 15  \times $ 
the number of vanishing labels $\Gm_j=0$ of $\mu$. 

\def\w{s}  
We claim that any such $\Gs$ with $\Gs_j=-1$ may be brought back to the fundamental Weyl chamber by a single Weyl reflection. 
To prove this point, take $\Gs=\sum_i \Gs_i \Go_i$ with all $\Gs_i\ge 0$ for $i\ne j$ and $\Gs_j=-1$. Then take 
the reflection $\w_j$ in the plane orthogonal to $\Ga_j$: 
$\w_j[\Go_i]=\Go_i -\Gd_{ij}\Ga_j= \Go_i -\Gd_{ij}\sum_{j'}\CC_{jj'}\Go_{j'}$,  with $\CC$ the Cartan matrix,
hence $\w_j[\Go_j]=-\Go_j + \sum_{j'\approx j} \Go_{j'}\,$, with the last sum running over 
the neighbours $j'$ of $j$ on the $E_6$ Dynkin diagram.
This gives (as $\Gs_j=-1$)
\bea
\w_j[\Gs]&=&\sum_{i\ne j} \Gs_i \Go_i+ \Go_j-\sum_{j'\approx j} \Go_{j'} \nonumber \\
&=&\Go_j+
\sum_{j'\approx j} (\Gs_{j'}-1) \Go_{j'}  +\sum_{i\ne j,\, i{\approx \!\!\!\!\slash}  j} \Gs_i \Go_i\,.
\label{wjsigma}
\eea 
By inspection, one checks that if  some $\Gl'$ of $[\Go_2]$ or $[\Go_4]$ has $\Gl'_j=-2$,
all the $\Gl'_{j'}$ for $j'\approx j$ are non-negative, thus 
$\Gs_{j'} -1=\Gl'_{j'}+\Gm_{j'}+\Gr_{j'}-1\ge 0$ for $j'\approx j$, and for the other $i\ne j,\ i \non j$
(neither $j$ nor one of its neighbours),   
  $\Gs_i=\Gl'_i +\Gm_i+\Gr_i \ge -1+0+1= 0 $, 
so that  all labels of $w_j[\Gs]$ in (\ref{wjsigma}) are non-negative, qed. 

$\bullet$ Among these $\CNr$ weights $\w_j[\Gs]$ that have been reflected, $\CNiid$ have a vanishing 
Dynkin label.   According to  (\ref{wjsigma}), 
  this may happen only (a) if $\Gl'_{j'}=0$ (and $\mu_{j'}=0$) for some $j'\approx j$, or  (b) if $\Gl'_i=-1$, $i\ne j,\ i\non j$ (and $\mu_i=0$).
\short{By inspection, one checks that for any node $j=1,\cdots, 6$   there exist 3 weights $\Gl'$ in $[\Go_2]$ or in $[\Go_4]$ such 
  that $\Gl'_j=-2$ and $\Gl'_{j'}=-1$ for each $j'$ ``neighbour" of $j$, thus 3 cases of type (a) per neighbour;
  and likewise one checks that there are four 
  $\Gl'\in [\Go_2]$ or   $\Gl'$ $\in [\Go_4]$ satisfying condition (b) for each pair of $(j,i)$ such that $\Gm_j=\Gm_i=0$.}
 Note that the fulfilment of these conditions is independent of the value of the non-vanishing labels of $\mu$. 
  There are, however, configurations where conditions (a) and/or (b) are
  satisfied for two pairs $(j,j')$ or $(j,i)$, see an example below,  and this depends 
  on the detailed location of the vanishing labels of $\mu$. We thus
  found more expedient to write a Mathematica$^{\rm \texttrademark}$ code to 
enumerate the $62=2^6-2$ configurations of vanishing labels of $\mu\ne 0$, and
  for each of them, to count the number of $\Gl'$ in $[\Go_2]$ or in $[\Go_4]$ that contribute to $\CNiid$. As 
  expected we found the same numbers for $[\Go_2]$ and $[\Go_4]$. We conclude that for a given
  $\mu$, the number $\CNiii$ of weights contributing negatively to the total multiplicity is the same
  for $[\Go_2]$ and $[\Go_4]$.

$\bullet$ We finally turn our attention to those  weights that need not be reflected. Their 
number $\CNnr= 351 -\CNr$  is the same for  $[\Go_2]$ and $[\Go_4]$. It remains to 
count the number $\CNiiu$ that lie on one of the walls of the fundamental Weyl chamber. 
There too, it is easy to see that there is only a finite number of cases to consider. 
Indeed if $\Gs$ has all its labels non-negative, $\Gs_j=0$ occurs if $\Gl'_j=-2$ and
$\Gm_j=1$, or  if $\Gs_j=-1$ and $\Gm_j=0$. There are $\sum_{\ell=1}^6 {6\choose \ell}2^\ell-1 =727$ choices 
for the   labels of $\mu$ equal to 0 or 1, and 
one may  write a code to check that for each of them, the number of $\Gl'$ leading to
a $\Gs$ on a wall is the same for $[\Go_2]$ and $[\Go_4]$.

We conclude that $\CNi=\CNnr -\CNiiu$ is the same for $[\Go_2]$ and $[\Go_4]$, and
so is $\sum_\nu N_{\Go_{2/4}\Gm}^{\qquad \nu} = \CNi-\CNiii$, thus completing the proof of our
assertion and of Lemma 1.  \hfill$\square$

\bigskip
{\bf An explicit example}\\
Let us illustrate the previous considerations on an explicit example.
Take the weight 
 $\mu = (1,0,0,0,2,0)$. In the tensor product with $\Go_2$, 
there are $\CNnr=351- 4\times 15= 291$ weights $\sigma=\lambda^\prime + \mu + \rho$
that have  non-negative labels  and thus 
belong to the fundamental chamber, and among them, 
$\CNi=38$ weights  that  do not belong to its walls. The corresponding  weights $\lambda^\prime + \mu$  give the following contribution to the tensor product  :
{\small
\begin{eqnarray*}
\!\!\!\!\!\!\!\!\!\!\!\!\!\!\!{} & & 5 (0, 0, 0, 0, 2, 0) + 
 5 (0, 0, 0, 0, 2, 1) + (0, 0, 0, 1, 0, 0) + (
  0, 0, 0, 1, 0, 1) + (0, 0, 1, 0, 2, 0)   \\
\!\!\!\!\!\!\!\!\!\!\!\!\!\!\!{} &  &+ 5 (0, 1, 0, 0, 1, 0) + (0, 1, 0, 0, 1, 1) + (
  1, 0, 0, 0, 0, 1) + 5 (1, 0, 0, 0, 3, 0) + 
 5 (1, 0, 0, 1, 1, 0)  \\
\!\!\!\!\!\!\!\!\!\!\!\!\!\!\! {} & &  + (1, 0, 1, 0, 0, 0) + (
  1, 1, 0, 0, 2, 0) + 
 5 (2, 0, 0, 0, 1, 0) + (2, 0, 0, 0, 1, 1)
\end{eqnarray*}
}
Among the $\CNr=15\times 4= 60$ weights $\sigma$ 
 that could lead to a situation of type (iii), 39 lie on a wall; \short{this 39 comes about in the following way:
 there are $3+3\times 3 +3 +3=18$ cases of type (a) in the discussion above, coming from a $\Gl'_j=-2$ on node
 $j=2,3,4, 6$, respectively, and $\Gl'_{j'}=0$ on a node $j'\approx j$ with $j'\in\{2,3,4,6\}$; and there are $4\times 6-3$
 cases of type (b), with 4 cases for each pair $(j,i)$ of non neighbours taken in $\{2,3,4,6\}$, but there is a 
 double counting of three weights that fulfil (b) for two such pairs, for example, $ (0, -1, 2, -1, 1, -2)$, whence the $-3$; 
 and $18+ 24-3=39$.}
 and  there are thus only $\CNiii=21$   that have no vanishing Dynkin label after reflection.
 For all these weights $\Gs$,  $w[\sigma] - \rho$ therefore gives a negative contribution to the tensor product. 
 One finds :
{\small
\begin{eqnarray*}
{} & &  4(0, 0, 0, 0, 2, 0) + 3(0, 0, 0, 0, 2, 1) +  3(0, 1, 0, 0, 1, 0) +  \\
{} & &4(1, 0, 0, 0, 3, 0) +  3(1, 0, 0, 1, 1, 0) + 4(2, 0, 0, 0, 1, 0)
\end{eqnarray*}
}
Substracting the second contribution from the first, one  gets the final result
{\small
\begin{eqnarray*}
\omega_2 \otimes \mu & = &
(0, 0, 0, 0, 2, 0) + 
 2 (0, 0, 0, 0, 2, 1) + (0, 0, 0, 1, 0, 0) + (
  0, 0, 0, 1, 0, 1) + \\
{} & &  
  (0, 0, 1, 0, 2, 0) + 
 2 (0, 1, 0, 0, 1, 0) + (0, 1, 0, 0, 1, 1) + (1, 0, 0, 0, 0, 1) +
 \\
 {} & &  
  (1, 0, 0, 0, 3, 0) +  2 (1, 0, 0, 1, 1, 0) + (1, 0, 1, 0, 0, 0) + (  1, 1, 0, 0, 2, 0) + \\
   {} & &  
   (2, 0, 0, 0, 1, 0) + (2, 0, 0, 0, 1, 1)\ .
\end{eqnarray*}
}
\ommit{
 In terms of dimensions, it reads 
{\small
\begin{eqnarray*}
[351] \otimes [7722] = \!\!\!\! &
\overline{[351]}^\prime +
 2 \overline{[19305]}+ \overline{[351]} + \overline{[17550]}+ \overline{[494208]}+ 
 2 [7371]+ \overline{[314496]}
  \\
 {}
+  [1728]  \!\!\!\!\!\!\!\!\!&+\overline{[61425]}+ 
 2 \overline{[112320]}+ [51975]+ [1123200]+ \overline{[7722]}+ [359424]
\end{eqnarray*}
}
where $[351]=\omega_2$,  $\overline{[351]} = \omega_4$ and $\overline{[351]}^\prime = (0, 0, 0, 0, 2, 0)$, 
which provides a welcome check of the calculation. }
\noindent
The total multiplicity is therefore $\CNi-\CNiii= 38 - 21=17$.

\smallskip

 If we now perform the same analysis for the tensor product $\omega_4 \otimes \mu$ with the same 
$\mu = (1,0,0,0,2,0)$, we again get a positive contribution of $38$ terms from the weights belonging to the fundamental chamber, and a negative contribution of $21$, from the reflected weights, so that the total multiplicity, $17$, is the same.
It may be noticed that the obtained weights for $\omega_2 \otimes \mu $ and $\omega_4 \otimes \mu$ are quite different, both for the two contributions and for their sum.
The final decomposition of  $\omega_4 \otimes \mu$ reads as follows:
 {\small
\begin{eqnarray*}
\omega_4 \otimes \mu 
& = & (0, 0, 0, 0, 3, 0) + (0, 0, 0, 0, 3, 1) + 2(0, 0, 0, 1,1, 0) + (0, 0, 0, 1, 1, 1)  + \\
{} & {} &  (0, 0, 1, 0, 0, 0) +  2(0, 1, 0, 0, 2, 0) +  (0, 1, 0, 1, 0, 0) + (1, 0, 0, 0, 1, 0) + \\
 {} & {} &  2(1, 0, 0, 0, 1, 1) + (1, 0, 0, 1, 2, 0) + (1, 0, 1, 0, 1, 0) + (1, 1, 0, 0, 0, 0) + \\ 
{} & {} &  (2, 0, 0, 0, 2, 0) + (2, 0, 0, 1, 0, 0)  
\end{eqnarray*}
}
\ommit{
or, in terms of dimensions,
{\small
\begin{eqnarray*}
\overline{[351]} \otimes [7722] =  \!\!\!\! &
\overline{[3003]}+ \overline{[146432]}+ 
 2 \overline{ [5824]}+ \overline{[252252]}+ [2925]+ 
 2 \overline{[78975]}+ [70070] \\
  {}  & \!\!\!\! 
+ [650]+ 
 2 [34749]+ \overline{[972972]}+ [852930]+ [5824]+ [85293]+ [78975]\,.
\end{eqnarray*}
}}

\subsection{Sums of fusion coefficients in $\hat E_6$}
 \label{appA2}

Let us now see how the presence of the back wall affects the  previous counting.
We have to examine what happens to weights $\Gs=\Gl'+\Gm+\Gr$ that are
either on or ``beyond" the shifted back wall, {\it i.e.} have  $\Gs_0\le 0$,
and we have to see under which condition  
some reflected weight may lie on a wall (and hence not contribute
to the multiplicity, according to  the affine Racah--Speiser algorithm).

a. First consider any weight $\Gs$ that undergoes a reflection as in sect.
\ref{appA1}. We
prove that $s_j[\Gs]$ lies within the shifted principal alcove
$ \CK(s_j[\Gs]) \le k+h^\vee$, {\it including its back wall}. Here and in the following, $j$ takes values
in $\{1,2,3,4,5,6\}$. \\
As in the previous section, we take  $\Gs$ with some $\Gs_j=-1$.  By
inspection, the weights $\Gl'$ that have $\Gl'_j=-2$  have level $\CK(\Gl')\le 1$, hence
$ \CK(\Gs)=\CK(\Gl')+\CK(\mu) +\CK(\Gr)\le 1 + k + (h^\vee-1)= k+h^\vee$, and
for $s_j[\Gs]=\Gs- \langle\Ga_j,\Gs\rangle \Ga_j= \Gs- \Gs_j \Ga_j= \Gs+\Ga_j$,
$ \CK(s_j[\Gs])= \CK(\Gs)+\CK(\Ga_j)$. The levels of the simple roots
$\Ga_j$ of $E_6$ are
$(0,0,0,0,0,1)$ for $j=1,\cdots, 6$. The previous inequality
gives us $\CK(s_j[\Gs])\le  k+h^\vee$ for $j=1,\cdots, 5$, while $s_6$,
with the root $\Ga_6$ of
level 1, looks more problematic. Fortunately, one checks by inspection
that all $\Gl'$ with
their 6-th Dynkin label equal to $-2$ have a level less or equal to 0,
(reducing the previous
bound by one unit), and thus we also have $\CK(s_6[\Gs])\le  k+h^\vee$.

b. We then turn to  cases where $\Gs$  is within the fundamental chamber but
``beyond" the shifted back wall, {\it i.e.}  has $\Gs_0 < 0$. Since
\be \label{Gsbw}  \Gs_0=k+h^\vee -\CK(\Gs)= (k-\CK(\mu)) +(1-\CK(\Gl'))\,, \ee
where the first bracket is non-negative, $\Gs_0<0$  occurs only for
$\CK(\Gl')=2$ and $\CK(\mu)=k$.
Such a $\Gs$ is brought back into the first (shifted) alcove by a single
(affine) Weyl reflection:
$s_0(\Gs)=\Gs -\theta$ whose \rung\ is indeed $\CK(s_0(\Gs))= k+h^\vee+1
-\CK(\theta)= k+h^\vee-1$ :
$ s_0[\Gs]-\Gr$ has \rung\ equal to $k$ and lies on the back wall of
$P_+^k$.
Now it is clear that the number of $\Gl'$  of level equal to 2 is the same in the
two (conjugate) weight systems $[\Go_2]$ and $[\Go_4]$.

c. We have finally to study cases where the  unreflected weight $\Gs$ or the
reflected $s[\Gs]$
lies on a wall of the alcove.
We leave aside the cases where the  weight $\Gs$ or $s_j[\Gs]$ lies on
one of the
ordinary walls of the Weyl chamber, that have been examined in the previous
subsection, and we focus on cases where $s_0[\Gs]$ is on one of the
ordinary walls, or  where $\Gs$ or $s_j[\Gs]$ is on the back wall.

In fact  it is not difficult  to show \\
-- that the number of unreflected $\Gs$ or reflected $s_j[\Gs]$ lying on
the back wall of the fundamental alcove
is the same for $[\Go_2]$ and $[\Go_4]$;\\
-- that the number of reflected $s_0[\Gs]$ lying on one or several
ordinary walls of the fundamental chamber
is the same for $[\Go_2]$ and $[\Go_4]$.\\
\short{This is proved as follows:\\
--  As shown by (\ref{Gsbw}),  $\Gs\in P_+$ is on the (shifted) back wall, {\it i.e.} $\Gs_0=0$,  iff 
$\mu$ is itself on the back wall ($\CK(\mu)=k$) and  $\Gl'$ is of level 1. It is clear 
that for such a $\Gm$, the number of such $\Gl'$ is the same in $[\Go_2]$ and $[\Go_4]$.\\
-- Consider cases where $\Gs$ has been reflected into $s_j[\Gs]$ which lies on the back 
wall of $P_+^{k+h^\vee}$. From the computation above, $\CK(s_j[\Gs])=\CK(\Gs)+\CK(\Ga_j)
=\CK(\Gs)+\Gd_{j6}$ which may be equal to $k+h^\vee$ only if $\CK(\Gs)=k$ and 
$\CK(\Gl')=1-\Gd_{j6}$. Now it is an easy matter to check that the number of $\Gl'$
such that $\Gl'_j=-2$ and $\CK(\Gl')=1-\Gd_{j6}$  is the same in $[\Go_2]$ and $[\Go_4]$.\\
-- Last we return to cases examined in b. above, where  $\Gs$ has to be reflected across the back wall, 
but assume now that  $s_0[\Gs]$ is on an ordinary wall of the fundamental chamber.
As seen above, $s_0[\Gs]=\Gs-\theta$, and $\theta=(0,0,0,0,0,1)$ in the basis of fundamental
weights, thus $(s_0[\Gs])_j =\Gs_j -\Gd_{j 6}$ which may vanish only for $j=6$ and $\Gs_6=1$ 
(we have assumed
that $\Gs$ was not on an ordinary wall, otherwise it would have dropped out in section A.1, hence $\Gs_j>0$).
It is clear that for any given $\Gm$, the number of $\Gl'$ such that $\Gl'_6=-\Gm_6$ 
 is the same in $[\Go_2]$ and $[\Go_4]$. qed }

The vanishing or negative contribution of these reflected weights to the
sum of fusion coefficients is thus the same for $\Go_2$ and $\Go_4$ and  we may finally
conclude that
$$ \sum_\Gn \hat N_{\Go_2\Gm}^{\quad \Gn} =\sum_\Gn \hat
N_{\Go_4\Gm}^{\quad \Gn}\,, 
$$
thus completing the proof of Theorem 2 for the $\hat E_6$ algebra.


{\bf An explicit example} (continuation)\\
The reader can illustrate the above discussion with the following example.
As in Appendix A1, we choose the irrep with highest weight $\mu=(1,0,0,0,2,0)$ that exists at levels $k \geq 3$.

{\small 

At level $k=3$, i.e. $q^{15}=-1$, \\
${}\qquad \omega_2 \otimes \mu = (0, 0, 0, 0, 2, 0)  + (0, 0, 0, 1,  0, 0) + (  0, 1, 0, 0, 1, 0) + ( 1, 0, 0, 0, 0,  1 )$; \\
quantum dimensions of the r.h.s. are ${\frac{1}{2} \left(5+\sqrt{5}\right),\frac{1}{2} \left(5+3 \sqrt{5}\right),\frac{1}{2} \left(5+3 \sqrt{5}\right),\frac{1}{2} \left(5+3 \sqrt{5}\right)}$. \\
${}\qquad \omega_4 \otimes \mu =  (0, 0, 0, 1, 1, 0) + (0, 0, 1, 0, 0, 0) +  (1, 0, 0, 0, 1, 0) + (1, 1, 0, 0, 0, 0)$;\\
quantum dimensions of the r.h.s. are ${2+\sqrt{5},\frac{3}{2} \left(1+\sqrt{5}\right),\frac{3}{2} \left(3+\sqrt{5}\right),2+\sqrt{5}}$.
\\
Total dimension is $(\frac{1}{2} \left(5+3 \sqrt{5}\right)) \times (\frac{1}{2} \left(5+\sqrt{5}\right)) = 5 \left(2+\sqrt{5}\right)$ in both cases, as it should.\\
Total multiplicity is $4$ in both cases.

At level $k=4$, i.e. $q^{16}=-1$\\
$\omega_2 \otimes \mu = (0, 0, 0, 0, 2, 0) + (0, 0, 0, 0, 4, 2)  +  (0, 0, 0, 1, 0, 0)  + (0, 0, 0, 1, 0, 1)  +  (0, 2, 0, 0, 2, 0)  +  (1, 0, 0, 0, 0, 1)  + (1, 0, 0, 0, 3,  0)  +  (2, 0, 0, 2, 2, 0)  +  (1, 0, 1, 0, 0, 0)  +  (2, 0, 0, 0, 1, 0)$\\
$\omega_4 \otimes \mu = (0, 0, 0, 0, 3, 0) + (0, 0, 0, 2, 2, 0) +  (0, 0, 1, 0, 0, 0) +  (0, 2, 0,0, 4, 0) +  (0, 1, 0, 1, 0, 0) +  (1, 0, 0, 0, 1, 0) +  (2, 0, 0, 0, 2,  2) +  (1, 1, 0, 0, 0, 0) +  (2, 0, 0, 0, 2, 0) +  (2, 0, 0, 1, 0, 0)$\\
The reader can check that both r.h.s. have total quantum dimension 
$(3+2 \sqrt{2}+\sqrt{2-\sqrt{2}}+3 \sqrt{2+\sqrt{2}}) \times (4+3 \sqrt{2}+2 \sqrt{2-\sqrt{2}}+4 \sqrt{2+\sqrt{2}}) = 52+37 \sqrt{2}+4 \sqrt{338+239 \sqrt{2}}$\\
 Total multiplicity is $10$ in both cases.
}

\section{Automorphisms of affine algebras}
\label{sec:automorphisms} 

We first describe these automorphisms for the algebras $\hat A_n$, $\hat D_{n=2s+1}$ and $\hat E_6$ which are used in section  \ref{sec:proofoftheorem2}   
for the proof of the vanishing of $\sum_\lambda S_{\lambda \kappa}$ when $\kappa$ is complex.
We then describe them for algebras $\hat B_n$, $\hat C_n$, $\hat D_{n=2s}$ and $\hat E_7$ which are used in section \ref{sec:quaternionic} 
that deals with the case where  $\kappa$ is quaternionic.
There are no non-trivial automorphisms for $\hat F_4$, $\hat G_2$ and $\hat E_8$.
These automorphisms reflecting the geometrical symmetries of the corresponding extended Dynkin diagrams are parametrized by the center of the chosen Lie group, or equivalently by classes of $P/Q$
where $P$ is the weight lattice, and $Q$ is the root lattice.  If $\ssigma$ is an automorphism of the Weyl alcove we have $S_{\ssigma(\lambda) \kappa} = \exp(\frac{2 \pi {i}  \tau(\kappa)}{N})   S_{\lambda, \kappa}$ where $\tau$ is the corresponding character of the center, and $N$ (sometimes called the connection index), the order of the center, is given by the determinant of the Cartan matrix.
Automorphisms of affine algebras are explicitly listed in \cite{DFMS} but the values of  $\tau$, the corresponding character of the center,  are not given  there.  The value of $\tau$ was calculated from the equality
  \be \exp[\frac{2  \pi i \tau(\kappa)}{N}] = \exp[-2 \pi {i}  \langle \kappa, \varpi \rangle]\ee
   where $ \langle \, , \,  \rangle$ is the canonical bilinear symmetric form of the root space, and where
$\varpi$ is an appropriate fundamental weight given as follows.
 Use the basic representation $\varpi = \omega_1$ for $A_n$, $B_n$, $E_6$,  $E_7$ and $\varpi = \omega_n$ for $C_n$. \colend
Use  $\varpi = \omega_n$ (one of the two spinorial irreps), $\varpi = \omega_{n-1}$ (the other spinorial) and $\varpi = \omega_{1}$, respectively for the three generators $\ssigma^{\prime}$, $\ssigma^{\prime\prime}$ and $\ssigma^{\prime\prime\prime}$ of the center $\Z_2 \times \Z_2$ of $D_{n=2s}$, each generator being equal to the product of the other two;
finally,  $\varpi = \omega_n$ (one of the two spinorial irreps) for the given generator of  $D_{n=2s+1}$. 
\bigskip
{\small
\noindent
\begin{center}
$
\begin{array}{c|c|c}
\goh_k & \textrm{center of } \go &  \textrm{$\CK(\Gl)$, generator(s), grading {and conjugate}:} \\
 & &  \\
\hline
& &  \\
&& \CK(\Gl)= \sum_{i=1}^{n}\Gl_i\qquad \Gl_0=k-\CK(\Gl) \\
\hat A_n & \Z_{n+1} & \ssigma(\lambda) =(\Gl_0,\Gl_1,\Gl_2,\cdots,\Gl_{n-1})\\
& & \tau(\lambda)=  \sum_{i=1}^{n} i \Gl_i \; \mod n+1  \\
& & {\Gl=(\Gl_1,\cdots, \Gl_n) \leftrightarrow \bar \Gl=(\Gl_n,\cdots,\Gl_1)} \\
 & &  \\
\hline
 & &  \\ 
&&   \CK(\Gl)=\lambda_1+2 \sum_{j=2}^{n-2}\lambda_j +\lambda_{n-1}+\lambda_n \qquad \Gl_0=k-\CK(\Gl)   \\
& & 
\ssigma(\lambda) =(\lambda_n, \lambda_{n-2}, \lambda_{n-3}, \cdots, \lambda_1, \Gl_0)\\
       \hat D_{n=2s+1}  & \Z_4 &
       \tau(\lambda)= 2{\displaystyle \sum_{j=1,\atop j \, odd}^{n-2}} \lambda_j + \lambda_{n-1} + 3 \lambda_n \mod 4 \;   \textrm{if}\;  n = 1 \mod 4   \\
&  & \tau(\lambda)= 2{\displaystyle \sum_{j=1,\atop j \, odd}^{n-2}} \lambda_j + 3 \lambda_{n-1} +  \lambda_n \mod 4 \;  \textrm{if}\;  n = 3 \mod 4   \\
& &{\Gl=(\Gl_1,\cdots,\Gl_{2s},\Gl_{2s+1}) \leftrightarrow \bar \Gl=(\Gl_1,\cdots,\Gl_{2s+1},\Gl_{2s}) } \\
&&\\
\hline
 & &  \\
&&\CK(\Gl)=\lambda_1+2\lambda_2+3\lambda_3+2\lambda_4+\lambda_5+2\lambda_6\qquad \Gl_0=k-\CK(\Gl) \\
\hat E_6  & \Z_3 & 
{ \ssigma(\lambda) =(\lambda_0, \lambda_6,\lambda_3,\lambda_2,\lambda_1,\lambda_4) } \\
                 &          & \tau(\lambda)=  2\lambda_1+\lambda_2+2\lambda_4+\lambda_5  \mod 3  \\
                 & & {\Gl=(\Gl_1,\Gl_2,\Gl_3,\Gl_4,\Gl_5, \Gl_6) \leftrightarrow \bar \Gl=(\Gl_5,\Gl_4,\Gl_3,\Gl_2,\Gl_1, \Gl_6)} \\
\hline
\end{array}
$
\end{center}
}

\smallskip

{\small
\begin{center}
$
\begin{array}{c|c|c}
\goh_k & \textrm{center of } \go &  \textrm{generator(s) and grading:} \\
 & &  \\
\hline
& &  \\
\hat B_n & \Z_{2} & \ssigma(\lambda) =(k - \Gl_1 - 2 \sum_{i=2}^{n-1}\Gl_i - \Gl_n,\Gl_2,\Gl_3,\cdots,\Gl_{n})\\
& & \tau(\lambda)=  \Gl_n \; \mod 2  \\
 & &  \\
\hline
 & &  \\
\hat C_n  & \Z_2 & { \ssigma(\lambda) = (\lambda_{n-1}, \lambda_{n-2}, \ldots, \lambda_2, \lambda_1, k - \sum_{i=1}^{n}  \lambda_i) } \\
                 &          & \tau(\lambda)=  \sum_{j \, odd} \;  \lambda_j   \; \mod 2  \\
& &  \\
\hline
 & &  \\
 \hat D_{n=2s}  & \Z_2 \times \Z_2& \ssigma^{\prime}(\lambda) = (\lambda_{n-1}, \lambda_{n-2}, \ldots, \lambda_2, \lambda_1, k - \lambda_1 - 2 \sum_{i=2}^{n-2} \lambda_i - \lambda_{n-1} - \lambda_n )\\
                               &         &  \tau^{\prime}(\lambda)= 2  \sum_{j=1, j \, odd}^{n-3} \lambda_j  + 2 \lambda_{n}   \mod 4   \\   
 &  & \\ 
 &  & \ssigma^{\prime\prime}(\lambda) =(\lambda_n, \lambda_{n-2}, \dots, \lambda_2,k - \lambda_1 - 2 \sum_{i=2}^{n-2} \lambda_i - \lambda_{n-1} - \lambda_n,  \lambda_{1} )\\
                               &         &  \tau^{\prime\prime}(\lambda)= 2  \sum_{j=1, j \, odd}^{n-3}  \lambda_j  + 2 \lambda_{n-1}  \mod 4   \\   
&  & \\ 
 &  & \ssigma^{\prime \prime\prime}(\lambda) =   (k - \lambda_1 - 2 \sum_{i=2}^{n-2} \lambda_i - \lambda_{n-1} - \lambda_n, \lambda_2, \lambda_3, \dots, \lambda_{n-2}, \lambda_n, \lambda_{n-1} ) \\
                               &         &  \tau^{\prime \prime\prime}(\lambda)= 2 \lambda_{n-1} + 2 \lambda_{n}   \mod 4   \\
& &  \\
\hline
 & &  \\
 \hat E_7  & \Z_2 & { \ssigma(\lambda) = (k-\lambda_1 - 2 \lambda_2 -3 \lambda_3 -4 \lambda_4 - 3\lambda_5 -2 \lambda_6 - 2 \lambda_7) } \\
                 &          & \tau(\lambda)=  \lambda_1+\lambda_3 + \lambda_7  \; \mod 2  \\
\hline
\end{array}
$
\end{center}
}

{\small
Conventions: $B_n$ has $n-1$ long simple roots, the last root $\alpha_n$ is short. $C_n$ has $n-1$ short simple  roots, the last root $\alpha_n$ is long. 
For $E_7$,  the root $\alpha_7$, at the extremity of the short branch is above the fourth vertex,  counted from the left (this is not the convention of \cite{DFMS}).}


\section{Types of representations for complex Lie groups and Lie algebras}
\label{sec:representationtypes}

\subsection{A collection of known results}
The following results are well known,  
see  for instance \cite{Malcev, Fell, Simon}, and are gathered here for the convenience of the reader.

{%
\small
\noindent
$
\begin{array}{c|c|c|c|c|}
\go & & complex & \multicolumn{2}{c}{self-conjugate} \\
\hline
{}  & {} & {}Ê& real & quaternionic \\
\hline
\hline
A_n & n = 1 \mod 4 &     \mu_j \neq \mu_{n+1-j} &  \multicolumn{2}{c|}{ \mu_j = \mu_{n+1-j}} \\
        &                          &      &   \mu_{(n+1)/2}   \;  \textrm{even}       &  \mu_{(n+1)/2} \;  \textrm{odd}  \\
        & & & & \\
\cdashline{2-5}
 & & & &\\
       & n \neq 1 \mod 4  &  \mu_j \neq \mu_{n+1-j} &  \multicolumn{2}{c|}{ \mu_j = \mu_{n+1-j}} \\
       &                             &   & \textrm{always} &\textrm{never}\\
\hline
B_n & n = 0,3 \mod 4 &      \textrm{never} &  \multicolumn{2}{c|}{\textrm{always} } \\
        &                          &      &   \textrm{always}     &  \textrm{never}   \\
        & & & & \\
\cdashline{2-5}
 & & & & \\
       & n =1,2 \mod 4  &   \textrm{never}   &  \multicolumn{2}{c|}{\textrm{always} } \\
       &                             &   & \mu_n \;  \textrm{even}  &  \mu_n \;  \textrm{odd} \\
\hline
C_n &   &      \textrm{never} &  \multicolumn{2}{c|}{\textrm{always} } \\
        &                          &      &  \mu_1 +\mu_3 + ... + \mu_m   \;  \textrm{even}  &  \mu_1 +\mu_3 + ... + \mu_m   \;  \textrm{odd} \\
&  &     &  \multicolumn{2}{c|}{m=n \, \textrm{if} \, n \, \textrm{is odd and}\,  m=n-1\, \textrm{if}\, n \,\textrm{is even} } \\
        & & & & \\
 \hline
D_n & n = 0 \mod 4 &   \textrm{never}  &  \multicolumn{2}{c|}{ \textrm{always} } \\
        &                          &      &    \textrm{always}       &  \textrm{never}  \\
        & & & & \\
\cdashline{2-5}
 & & & & \\
       & n =2  \mod 4  &  \textrm{never}    &  \multicolumn{2}{c|}{\textrm{always} } \\
       &                             &   & \mu_{n-1}+\mu_n \; \textrm{even} & \mu_{n-1}+\mu_n \;  \textrm{odd} \\
       \cdashline{2-5}
 & & & &\\
       & n = 1, 3  \mod 4  &  \mu_{n-1} \neq \mu_n   &  \multicolumn{2}{c|}        { \mu_{n-1} = \mu_n}  \\
       &                             &   &  \textrm{always}  & \textrm{never}  \\
\hline
E_6  &  &    \mu_1 \neq \mu_5 \; \textrm{or}\;  \mu_2 \neq \mu_4 &  \multicolumn{2}{c|}{ \mu_1=\mu_5 \; \textrm{and}\;  \mu_2=\mu_4 } \\
        &                          &      &   \textrm{always}  & \textrm{never}  \\
        & & & & \\
\hline
E_7  &  & \textrm{never}  &  \multicolumn{2}{c|}{\textrm{always} } \\
        &                          &     & \mu_1+\mu_3+\mu_7  \;  \textrm{even}  &\mu_1+\mu_3+\mu_7  \;  \textrm{odd} \\
        & & & & \\
        \hline
E_8  &  & \textrm{never}  &  \multicolumn{2}{c|}{\textrm{always} } \\
        &                          &     &  \textrm{always}  &  \textrm{never} \\
        & & & & \\
        \hline 
G_2  &  & \textrm{never}  &  \multicolumn{2}{c|}{\textrm{always} } \\
        &                          &     &  \textrm{always}  &  \textrm{never} \\
        & & & & \\
        \hline          
F_4  &  & \textrm{never}  &  \multicolumn{2}{c|}{\textrm{always} } \\
        &                          &     &  \textrm{always}  &  \textrm{never} \\
        & & & & \\
        \hline                
\end{array}
$
}


\subsection{Fusion and the Frobenius-Schur indicator}
\label{sec:FS}
According to \cite{Bantay}, see also \cite{SchauNg}, the second indicator $I_\mu$ of  Frobenius-Schur, whose value is $1, 0$ or $-1$, according to the type (real, complex or quaternionic) of the representation $\mu$ of $\go$, can be obtained as
$$
I_\mu = \sum_{\nu \sigma} S_{0 \sigma} \,  \hat N_{\mu\nu}^{\ \ \sigma} \, S_{0 \nu} \, \frac{\iota(\sigma)^2}{ \iota(\nu)^2}
$$
where $\iota(\nu)=
\exp( 2 i \pi {\frak h}(\nu))$ and  ${\frak h}(\nu) = {\langle \nu, \nu+2 \rho\rangle}/{(\langle \theta, \theta \rangle ( k+h^\vee))}$ is the conformal weight of $\nu$.\\
\noindent
It is not too difficult to show that $\iota(\nu) = T_{\nu\nu} \; \psi$  where
 $\psi = \exp(2 i \pi c  /24)$,  
 $c = {\dim(\go) \, k}/{(k+h^\vee)}$ is the central charge, 
and $T$ is the modular matrix that obeys, together with $S$, the usual relations $(ST)^3 = S^4=1$. 
 Using the fusion matrix $\hat N_\mu$,
the previous relation between $\iota$ and $T$, the fact that $S$ is symmetric, $T$ is diagonal, $S^{-1}= S\, C$ and that $C_{0p}= \delta_{0p}$, we recast the formula giving the indicator as follows:
$$
I_\mu =   {(S^{-1} T T \, \hat N_\mu \, T^{-1} T^{-1} S)}_{00}
$$
This last expression can be used to check easily the type of representations discussed in the text.

{\bf Acknowledgements} 

We wish to thank P. Di Francesco,  V. Petkova and especially P. Dorey  for valuable discussions.\\
Part of this paper was written while one of us (R.C.) was a visiting professor at the University of Tokyo, and at RIMS (Kyoto).  Their hospitality and support are gratefully acknowledged.

Partial support from ANR program ``GRANMA'' 
ANR-08-BLAN-0311-01 
and from ESF ITGP network is also acknowledged. 


\end{document}